\documentclass[iop]{emulateapj}
\usepackage{amsmath}

\begin{document}
\title{Synthesizing Exoplanet Demographics from Radial Velocity and Microlensing Surveys, I: Methodology}  

\author{Christian Clanton,
B. Scott Gaudi}
\affil{Department of Astronomy, The Ohio State University, 140 W. 18th Ave., Columbus, OH 43210, USA}
\email{clanton@astronomy.ohio-state.edu}

\begin{abstract}
Motivated by the order-of-magnitude difference in the frequency of giant planets orbiting M dwarfs inferred by microlensing and radial velocity (RV) surveys, we present a method for comparing the statistical constraints on exoplanet demographics inferred from these methods. We first derive the mapping from the observable parameters of a microlensing-detected planet to those of an analogous planet orbiting an RV-monitored star. Using this mapping, we predict the distribution of RV observables for the planet population inferred from microlensing surveys, taking care to adopt reasonable priors for, and properly marginalize over, the unknown physical parameters of microlensing-detected systems. Finally, we use simple estimates of the detection limits for a fiducial RV survey to predict the number and properties of analogs of the microlensing planet population such an RV survey should detect. We find that RV and microlensing surveys have some overlap, specifically for super-Jupiter mass planets ($m_p \gtrsim 1~M_{\rm Jup}$) with periods between $\sim 3-10~$years. However, the steeply falling planetary mass function inferred from microlensing implies that, in this region of overlap, RV surveys should infer a much smaller frequency than the overall giant planet frequency ($m_p\gtrsim 0.1~M_{\rm Jup}$) inferred by microlensing. Our analysis demonstrates that it is possible to statistically compare and synthesize data sets from multiple exoplanet detection techniques in order to infer exoplanet demographics over wider regions of parameter space than are accessible to individual methods. In a companion paper, we apply our methodology to several representative microlensing and RV surveys to derive the frequency of planets around M dwarfs with orbits of $\lesssim 30~$years.
\end{abstract}

\keywords{methods: statistical -- planets and satellites: detection -- planets and satellites: gaseous planets -- techniques: radial velocities -- gravitational lensing: micro -- stars: low-mass}

\section{Introduction}
\label{sec:introduction}
The field of exoplanets has reached the state where statistically significant samples of planets have been detected via multiple different methods: radial velocities, transits, microlensing, and direct imaging. Each of these techniques are sensitive to different, yet complementary regions of parameter space, both in terms of the orbits and physical properties of planets, as well as their host stars. This means that it should be possible to derive generalized constraints on the demographics of planets over a broader region of parameter space than is available to each method individually by synthesizing their various data sets. This is not a trivial task, as each method of exoplanet discovery introduces unique observational biases and selection effects that have to be reconciled before comparing directly with detections from other methods.

Various RV and microlensing surveys have provided independent constraints on the demographics of exoplanets \citep{2005ApJ...622.1102F,2008PASP..120..531C,2008A&A...487..373S,2009A&A...493..639M,2010Sci...330..653H,2010PASP..122..905J,2010ApJ...720.1073G,2010ApJ...710.1641S,2011arXiv1109.2497M,2013A&A...549A.109B,2011Natur.473..349S,2012Natur.481..167C}. These two methods are generally complementary. RV surveys are sensitive to low-mass planets at short periods and larger planets at longer periods (ultimately limited to planets with periods less than the duration of the survey), while microlensing is sensitive to planets with projected separations near the Einstein radius of their host stars, which generally corresponds to larger separations than are covered by RV surveys ($\sim 4~$AU for a Solar mass star, and scaling as the square-root of the mass of the host star).

With regards to giant planets, this complementarity of RV and microlensing surveys is important. The two techniques are generally probing regions of parameter space where we expect planet formation to operate differently. RVs are most sensitive to orbits interior to the snow line, whereas microlensing is more sensitive to orbits exterior to the snow line. The snow line is the point in protoplanetary disks at (and beyond) which water ices can deposit out of the nebular gas \citep{1981PThPS..70...35H}. Past the snow line, deposition of water ices increases the surface density of the disk by a factor of $\sim 2-3$, increasing the isolation mass by a factor of $\sim 4-5$, allowing for gas giant formation by core accretion within the observed, short lifetimes ($\sim 1-10~$Myr) of gaseous disks \citep{2008ApJ...673..502K,1995Natur.373..494Z,2006ApJ...651.1177P}, at least for Solar-mass stars. Giant planets detected by RV are thus generally expected to have migrated while those detected by microlensing presumably formed \emph{in situ}. Therefore, by combining constraints on exoplanet demographics inferred by the two methods, we can get a complete picture of giant planet formation and migration, and disentangle these two competing physical effects.

RV and microlensing surveys also differ in terms of host stars: RV samples are targeted, but this is not the case for microlensing. The rarity and unpredictability of microlensing events render a targeted survey ineffective, so the general observing strategy is to instead monitor large numbers of stars in the most dense fields towards the Galactic bulge. Such serendipitous microlensing events are dominated by M dwarf lenses, as these are more common than higher mass lenses.
  
In terms of giant planets, which we define as having masses $m_p \gtrsim 0.1~M_{\rm Jup}$ (see the companion paper, \citet{clanton_gaudi14b}, for a discussion of what exactly constitutes a ``giant planet''), comparing detection results from RV and microlensing surveys is particularly interesting, because giant planets formed by core accretion are predicted to be rare around M dwarfs \citep{2004ApJ...612L..73L}. This prediction has been claimed to be confirmed by RV surveys of M dwarfs, who find a paucity of giant planets \citep{2010PASP..122..905J,2013A&A...549A.109B} relative to Solar-mass stars. However, because these surveys are mostly sensitive to separations less than the snow line, it is not clear if this is a consequence of a lack of formation or a lack of migration. Indeed, microlensing surveys \citep{2010ApJ...720.1073G,2012Natur.481..167C} find that giant planets beyond the snow line are more common by an order of magnitude than short period giant planets probed RV surveys, which on its face is in contradiction to the generic prediction of core accretion.

Motivated by this order of magnitude difference in giant planet frequency, here we develop the formalism and methodology for comparing and synthesizing the exoplanet detection results from RV and microlensing surveys. In a companion paper, we employ this methodology to determine if these two independent measurements of giant planet frequencies are consistent with the same population of planets.

We provide a high-level overview of the RV and microlensing techniques in \S~\ref{sec:mlens_rv_techniques}, highlighting the key differences necessary to understand before comparing exoplanet demographics derived independently from these two methods. In \S~\ref{sec:OoM} we provide an order of magnitude estimate that demonstrates that the ``typical'' microlensing planet, defined such that its parameters place it in the peak region of sensitivity for microlensing surveys, should be marginally detectable by RVs. We derive distributions of the period and velocity semi-amplitude expected from this ``typical'' microlensing planet, allowing for circular and eccentric orbits in \S~\ref{sec:typical_dists}. In \S~\ref{sec:marg} we develop a procedure to marginalize over all microlensing parameters, including the projected separation, the host star/planet mass ratio, host star mass, and lens distances to produce a posterior distribution of radial velocity observables for a population of planets analogous to that inferred from microlensing. We show this marginalized distribution and compute the number of planets per star a RV survey of M dwarfs (with some fiducial sensitivity and number of epochs per star) should detect, as well as the number of long-term RV trends per star resulting from planets, in \S~\ref{sec:results} and provide a discussion of our conclusions in \S~\ref{sec:discussion}.

\section{Microlensing and Radial Velocity Techniques}
\label{sec:mlens_rv_techniques}
The observables, and the physical processes that produce them, of RV and microlensing surveys are fundamentally different. The manner in which the targets are ``selected'' and the ways in which the two types of surveys are carried out are thus also different. This section provides a brief overview of both RV and microlensing techinques, and is meant to highlight key differences that are critical to understanding the comparison of exoplanet demographics derived independently from these two methods. For additional information on the methodology of RV planet searches and more complete reviews of the microlensing technique, see \citet{2004MNRAS.354.1165C}, \citet{2008PASP..120..531C}, and \citet{gaudi10,2012ARA&A..50..411G}.

{\bf Radial Velocities:} RV surveys monitor the time variance of the Doppler shift of absorption lines in stellar spectra, looking for variance resulting from orbital motion of the stars due to a companion (or companions). The primary measurements that are derived from such observations are the velocity semi-amplitude, $K$, and the period of variation, $P$. A measurement or a constraint on the eccentricity, $e$, of the system along with $K$ and $P$ yield a calculation of the mass function of the system, which, in the limit $m_p \ll M_{\star}$, has the form $(m_p\sin{i})^3/M_{\star}^2$. Here, $m_p$ is the mass of the planet, $i$ is the inclination (defined as the angle between the plane of the sky and the planet's orbital plane, such that $i=0^{\circ}$ corresponds to viewing the system face-on and $i=90^{\circ}$ corresponds to edge-on), and $M_{\star}$ is the mass of the host star. Combined with an estimate of $M_{\star}$, it becomes possible to compute the minimum planet mass, $m_p\sin{i}$, and the semimajor axis of the planet, $a$.

Practically, to detect a planet with RV, the measurement uncertainties (both systematic and intrinsic errors, such as those resulting from stellar jitter) must be low enough to distinguish the periodic variation of the Doppler signal. Ignoring instrumental errors, the uncertainty of an RV measurement of a typical stellar absorption line scales as $\sigma_{\rm RV} \propto \Delta^{3/2}/ (WI_0^{1/2})$, where $\Delta$ is the full-width at half-maximum of an absorption line, $W$ is its equivalent width, and $I_0$ is the continuum intensity (Beatty \& Gaudi, in preparation). This sets some general target requirements for RV surveys. In order to maximize sensitivity, it is best to observe stars with lots of absorption lines (spectral types FGKM), that emit lots of photons (bright, nearby), that are not rapidly rotating (typically no later than late F stars), and that are stable (not active, have few star spots, etc.).

The required observational cadence is such that sufficient coverage over all phases of RV variation is obtained. However, because the RV variation is periodic, this is not a strict requirement (although it does affect the overall sensitivity of a survey of a given duration). For instance, in order to detect a planet of period $P$, a RV survey must run for a duration (i.e. maximum time baseline) equal to at least $P$. In this case, the cadence must be significantly shorter than $P$ to obtain good phase coverage. On the other hand, if the duration of the survey is several times longer than $P$, then the cadence need not be so short. Though, in general, the shorter the cadence, the greater the period sensitivity of the survey.

Consequently, the signal-to-noise ratio (SNR) at which a RV survey can detect planets with a given period depends on the strength of the signal $K$, the magnitude of the measurement uncertainties $\sigma$, the duration of the survey $T$, and the total number of observations $N$. The time-dependent radial velocity of a star orbited by a planet on a circular orbit with period $P$ has the form $v\left(t\right) = K\sin{\left(2\pi t/P - \phi_0\right)}$, where $\phi_0$ is an arbitrary phase. If we assume a uniform and continuous sampling of the radial velocity curve, wherein observations at $N$ epochs are taken over a survey duration $T$, each with a measurement uncertainty $\sigma$, then we can estimate the SNR of the detection to be
\begin{equation}
{\rm SNR}\left(\phi_0\right) = \left\{\frac{N}{\sigma^2}\frac{1}{T}\displaystyle \int_{-T/2}^{T/2}\left[v\left(t\right) - \left<v\left(t\right)\right>\right]^2dt\right\}^{1/2}\; , \label{eqn:sn_phi0}
\end{equation}
where $\left<v\left(t\right)\right>$ is the time average of $v\left(t\right)$ and is a function of $\phi_0$. Averaging equation (\ref{eqn:sn_phi0}) over $\phi_0$, we obtain a ``phase-averaged'' SNR, which we designate by the variable $\mathcal{Q}$, for a RV detection:
\begin{align}
    \mathcal{Q} \equiv \left<{\rm SNR}\right>_{\phi_0} & {} = \left(\frac{N}{2}\right)^{1/2}\left(\frac{K}{\sigma}\right) \nonumber \\
    & {} ~\times \left\{1-\frac{1}{\pi^2}\left(\frac{P}{T}\right)^2\sin^2{\left(\frac{\pi T}{P}\right)}\right\}^{1/2}\; . \label{eqn:snr_full}
\end{align} 
When $\frac{P}{T} << 1$, the SNR for a RV detection is very nearly independent of period,
\begin{equation}
    \mathcal{Q} \sim \left(\frac{N}{2}\right)^{1/2}\left(\frac{K}{\sigma}\right)\; . \label{eqn:snr_small_p}
\end{equation}

As one might expect, more epochs, decreased measurement errors, and longer survey durations increase the SNR of a RV planet detection. For periods of roughly the survey duration ($P\sim T$), the actual behavior shown by equation (\ref{eqn:snr_full}) is slightly more complicated, but equation (\ref{eqn:snr_small_p}) is a good approximation. For the purposes of this paper, we assume that an RV survey must collect data over a time baseline $T \geq P$ in order to detect a planet of period $P$, however we assume no detection dependence on orbital eccentricity. In practice, planets on orbits with eccentricities $e\gtrsim0.6$ are much more difficult to detect by RV \citep{2004MNRAS.354.1165C}, but as we will show in later sections (see \S~\ref{subsec:eccentric_orbits} and \S~\ref{subsec:estimate_percent_mlens_detected}), we choose an eccentricity prior that puts a small fraction of planets on such eccentric orbits and thus our uncertainties due to this assumption are small compared to other sources of error.

RV observations rely on detecting light from host stars, requiring many photons to obtain high signal-to-noise ratio spectra. This can be advantageous, as it is often possible to obtain a wealth of information about RV-discovered planet hosting stars, including their colors, magnitudes, surface gravities, parallaxes, proper motions, and more. The trade off being that RV surveys are limited to the more luminous, or nearby, host stars. The basic observing strategy for most RV surveys is then to select a specific set of targets and monitor those targets for their duration. In contrast, microlensing surveys do not even require the collection of photons from host stars. 

{\bf Microlensing:} The microlensing technique monitors the magnification of a background star (i.e. the source) as it passes behind (or very nearly behind) a lensing system. Physically, microlensing probes the instantaneous gravitational structure of the lensing system, which determines the magnification pattern. The magnification pattern, together with the source trajectory and impact parameter, $u_0$, determines the shape of the observed light curve. Light curves resulting from a microlensing event produced by a binary acting as a gravitational lens have distinct features that reveal the binary nature of the lens, even at small mass ratios \citep{1991ApJ...374L..37M}. A planet orbiting a star is simply a binary lens system with a small mass ratio, $q \ll 1$, and will produce a magnification pattern, and thus light curve, that is dominated by the host star \citep{1999A&A...349..108D}. The planet can only detected when the source crosses a perturbation produced by the planet in the magnification pattern of the host star. The magnitude and form of the resultant planetary signal are determined by the size, shape, and location of these perturbations. Perturbations with a larger area on the sky have a higher probability of a source crossing, thus detection probabilities are higher for physically larger perturbations. The properties of the planetary perturbations are largely determined by the mass ratio $q$ and the projected separation $s$ in units of the angular Einstein radius, $\theta_E$.

There are five parameters routinely measured from a microlensing event due to a simple lens consisting of a single, isolated mass. The light curve alone allows one to measure the blend flux, $F_b$, the magnified source flux, $F_s$, the Einstein crossing time, $t_{\rm E}$, the impact parameter in units of the Einstein radius, $u_0$, and the time of maximum magnification, $t_0$. For a microlensing event due to a binary lens, consisting of a host star and planet, it is also typically possible to measure the projected separation in units of the Einstein radius, $s$, and the planet-to-star mass ratio, $q$, although in certain limiting cases, both $q$ and $s$ are subject to discrete and continuous degeneracies \citep{1999A&A...349..108D,1997ApJ...486...85G,1998ApJ...506..533G}. It is often not possible to measure the mass of the lens star, $M_l$, or the distance to the lens, $D_l$; to do so requires measurement of higher-order effects in the observed light curves, namely finite source effects and microlens parallax \citep{1966MNRAS.134..315R,1992ApJ...392..442G,1994ApJ...421L..75G,1994ApJ...430..505W,1995ApJ...449...42W,1996ApJ...471...64H}. It is also possible to infer $M_l$ and $D_l$ with measurements of the lens flux and finite source effects \citep{2007ApJ...660..781B}. Without measurements of $M_l$ and $D_l$, it is impossible to calculate the Einstein radius in physical units, $R_E$. Thus, in such cases, it is not possible to directly measure the planetary mass or projected separation in physical units.

There are multiple different channels through which planets can be detected by microlensing, each corresponding to different lens geometries and/or source trajectories:

\begin{itemize}
\item Source crossing of a planetary caustic resulting from either close or wide separation planets in low magnification events. Although such perturbations are unpredictable, this is referred to as the ``main'' channel for planet detections by microlensing because the planetary caustics have the largest area on the sky and thus have the highest probability of perturbing a source image magnified by the primary lens mass (i.e. the host star).
\item Perturbations from the central caustic resulting from either close or wide planets in high-magnification events. High-magnification events are relatively rare, but can be recognized in advance of the peak.
\item Resonant caustic perturbation in modest to high magnification events.
\item Isolated, short timescale events resulting from wide-separation or free-floating planets.
\end{itemize}

Another way of understanding these detection channels is to realize that planets are only detected when they perturb the images created by the primary lens (i.e. the host star).

The optical depth to a microlensing event in lines-of-sight towards the Galactic bulge is of order $10^{-6}$; the chance that a given star will lens a background source is small. A small fraction of stars that do undergo microlensing events will host planets that perturb the magnification pattern significantly. In a microlensing planet search, many stars must be monitored at a relatively high cadence. The typical (i.e. median) timescale of primary microlensing events is $\sim 20~$days, requiring a cadence of just a few days to detect. The timescale of planetary perturbations scales as $\sim q^{1/2}t_E$, which is roughly $\sim 15~$hours for a Jupiter-mass planet or $\sim 1~$hour for an Earth-mass planet orbiting a Solar-type star. Thus, the required cadence to accurately detect and characterize planetary systems needs to be of order tens of minutes or less.

Monitoring enough stars at a high enough cadence to detect planetary perturbations has not been possible until the relatively recent advent of large format CCD cameras with fields-of-view (FOV) of several square degrees. Rather, first generation microlensing planet searches have operated on the two-stage process advocated by \citet{1992ApJ...396..104G}, wherein survey collaborations such as the Optical Gravitational Lensing Experiment \citep[OGLE;][]{2003AcA....53..291U} and Microlensing Observations in Astrophysics \citep[MOA;][]{2008ExA....22...51S}, with large FOV cameras monitor several tens of square degrees at a cadence of a few observations a day or less. They reduce their data in real-time, alerting possible microlensing events typically before peak magnification. These alerts prompt follow-up teams, such as the Probing Lensing Anomalies NETwork \citep[PLANET;][]{1998ApJ...509..687A} and the Microlensing Follow Up Network \citep[$\mu$FUN;][]{2010ApJ...720.1073G}, that have access to narrow-angle telescopes with full longitudinal coverage in the Southern Hemisphere to monitor ongoing microlensing events on hour timescales.

The PLANET collaboration has substantial access to $0.6-1.5~$m telescopes located in South Africa, Perth, and Tasmania, allowing them to monitor dozens of events each season and search for planets via the main channel. The $\mu$FUN collaboration employs a single 1.3-m telescope in Chile to monitor promising alerted events in order to try to identify high-magnification events substantially before peak. When events are identified as high-magnification, other telescopes in the collaboration are engaged to obtain continuous coverage of the light curve during the high-magnification peak, which often gets as bright as $I \lesssim 15$. The high-magnification enables the use of relatively small-apertures ($0.3-0.4~$m), allowing amateur astronomers to contribute to the photometric follow-up, which make up over half of the members of the $\mu$FUN collaboration. This approach searches for planets by looking for perturbations from the central caustic, rather than the main channel.

{\bf Key Differences:} The main difference between observing strategies is that stars are targeted in RV surveys, while they not targeted in microlensing surveys. This is a reflection of the fact that these two techniques probe completely different physical mechanisms in the systems they observe. RV surveys search for periodic variations, and the information about a planetary system is contained within this periodicity. On the contrary, the information about a planetary system detected by microlensing is contained within a single light curve, the entirety of which is observed on $\sim $ month timescales. Microlensing has a unique advantage in that the technique is sensitive to planets on long-period orbits ($\sim $ tens of years) without a need to observe for comparable timescales.

Microlensing events are rare, unpredictable, and transient signals, resulting from a chance alignment of stars. The required alignment is very precise, and so the microlensing event rate per star is low. For a range of lens distances, $dD_l$, the contribution to the event rate scales as $\propto n\left(D_l\right)M_l^{1/2}$, where $n\left(D_l\right)$ is the number density of lenses and $M_l$ is the lens mass and thus the integrated event rate is explicitly dependent on the mass function of lenses \citep[see \S~\ref{subsec:event_rate} for more on event rates; see also][]{1991ApJ...366..412G,1994ApJ...430L.101K}. Consequently, microlensing surveys do not get to ``choose'' their targets; their targets are instead determined by the shape of the mass function of lenses \citep{2000ApJ...535..928G}.

RVs provide strong constraints on the properties of host stars and their planets, yielding measurements of $m_p\sin{i}$, $a$, and usually $e$. In a majority of cases, microlensing provides only $q$ and $s$, neither of which holds any information about orbital eccentricity or orientation. In addition, the projected separation is only weakly correlated with the orbital semimajor axis.

Another important distinction between the two techniques is their sensitivities to different phase spaces of planetary orbits. A host star will appear to have no radial velocity when viewed face-on ($i=0^{\circ}$), and will have maximum radial velocity when viewed edge-on ($i=90^{\circ}$). RV surveys are thus most sensitive to edge-on systems. In a microlensing event, bound planets are detected when they perturb the images created by the primary lens. In events where the source is significantly magnified, the images are always located near the Einstein ring of the primary. The detection probability for planets is thus maximized when the planet has a projected separation equal to the Einstein ring, $s = 1$. Microlensing surveys are most sensitive to face-on systems because there is more overlap between the (projected) planetary orbit and the Einstein ring of the primary than there is in edge-on systems.

The Einstein radius is physical units, $R_E = D_l\theta_E$, is given by
\begin{equation}
    R_E\left(M_l, D_l, D_s\right) = \sqrt{\frac{4GM_l}{c^2}\frac{D_l\left(D_s-D_l\right)}{D_s}} \label{eqn:einstein_ring_radius} \; ,
\end{equation}
where $D_l$ is the distance to the lens mass, $M_l$, and $D_s$ is the distance to the source. A lens with mass $0.5~M_{\odot}$ located at a distance $D_l=4~$kpc that is lensing a source located $D_s = 8~$kpc away has an Einstein radius $R_E \approx 2.9~$AU. Since the detection probability for a planet in a microlensing event is maximized when the projected separation is equal to the Einstein ring, microlensing has a peak sensitivity to planets located beyond the ice line, a distance of about $a_{\rm ice}= 1.7~{\rm AU}(M_l/0.5~M_{\odot})^{2/3}$ \citep{2008ApJ...673..502K}. As we have shown above, the sensitivity of RV surveys declines towards longer periods. However, this does not mean there is no overlap at all between RV and microlensing surveys.

A planet with mass $1~M_{\rm Jup}$ and a semimajor axis of $2.9~$AU orbiting a $0.5~M_{\odot}$ star on a circular orbit will produce a velocity semi-amplitude of $K \sim 25~{\rm m~s^{-1}}\left(P/7~{\rm yr}\right)^{-1/3}\left(m_p/M_{\rm Jup}\right)$ when viewed edge-on. For current state-of-the-art RV surveys, sensitivities of $\sigma\sim 1-2~{\rm m~s^{-1}}$ are routinely achieved, so detecting such a planet would simply be a matter of observing the target for at least $\sim 5~$years. Indeed, several such planets have been detected \citep{2013A&A...549A.109B,2014ApJ...781...28M}.

\section{Order of Magnitude Estimate of the Period and Velocity of a Typical Microlensing Planet}
\label{sec:OoM}
The main goal of this study is to develop a methodology for mapping the parameter space of analogs to the planets detected by microlensing into RV observables. Furthermore, we aim to gain a deep understanding of how survey sensitivities from both methods are imprinted on the distributions of RV observables we derive. Microlensing is sensitive to a given mass ratio, $q$, and projected separation, $s$. With $s$ and the Einstein radius, $R_E$, we can calculate the projected separation in physical units, $r_{\perp} = sR_E$. The combination of $q$ and the lens mass, $M_l$, yields the planet mass, $m_p$. We can then map $r_{\perp}$ into a semimajor axis, $a$, and thus a period, $P$, which together with $m_p$, maps into a velocity semi-amplitude, $K$. Of course, there are unknown physical parameters of microlensing detected systems\footnote{Technically, the exact source distance, $D_s$, is also unknown, although the vast majority of sources are in the Galactic bulge, and therefore there exists a strong prior on $D_s$.}, namely $M_l$ and the distance to the lens, $D_l$. Thus, in mapping a given microlensing detection, characterized by a $\left(q,s\right)$ pair, we must adopt reasonable priors on $M_l$ and $D_l$ and marginalize over them to infer the true distribution of $\left(r_{\perp}, m_p\right)$. Since a microlensing detection offers no constraints on the orientation of a planet's orbit, priors on the orbital parameters must be adopted to complete the final step of mapping into a $\left(K,P\right)$.

In this section, we begin by providing an order of magnitude estimate of the radial velocity observables $\left(K,P\right)$ that would be produced by a ``typical'' microlensing planet on a circular orbit. We define the ``typical'' microlensing planet as one residing in the peak region of sensitivity with the following parameters: a host star mass of $M_l \sim 0.5~M_{\odot}$, a mass ratio of $q \sim 5\times 10^{-4}$, and a projected separation of $r_{\perp} \sim R_E\sim 2.5$~AU, where $R_E \sim 2.5~$AU is the typical Einstein radius for a $0.5~M_{\odot}$ lens \citep{2010ApJ...720.1073G}. This corresponds to a typical microlensing planet mass of $m_p \sim qM_l\sim 0.26~M_{\rm Jup}$~$\sim~$M$_{\rm Sat}$.

The semi-major axis of a planet in terms of its projected separation is given by 
\begin{equation}
    a = \frac{r_{\perp}}{\sin{\beta}}\; ,
\end{equation}
where $\sin{\beta} = \sqrt{\cos^2\theta + \sin^2\theta\cos^2i}$, $\theta$ is the orbital phase angle and $i$ is the inclination of the planetary orbit. For this order of magnitude estimate, we assume a median value of the $\sin{\beta}$ distribution. For randomly-oriented, circular orbits, $\theta$ will be distributed uniformly as $0 \leq \theta < 2\pi$, and the distribution of inclinations is such that $\cos i$ is uniformly distributed as $0 \leq \cos i \leq 1$. The median value of $\cos i$ is simply 0.5, so the median value of $\sin i$ is given by $\sqrt{1-\left(0.5\right)^{2}} \approx 0.866$. The distribution of $\sin \beta$ is identical to the distribution of $\sin i$ in the case of circular orbits, so their median values are exactly the same. This gives us that the typical semi-major axis for a microlensing planet is $a_{\rm typ} \approx \frac{r_{\perp}}{0.866}$. We can now compute the orbital period from Kepler's 3rd law:
\begin{align}
    P_{\rm typ} \approx & {} \frac{2\pi}{\left(GM_l\right)^{1/2}}a_{\rm typ}^{3/2} \nonumber \\
        \approx & {} 7~{\rm yr}\left(\frac{M_l}{\rm 0.5~M_{\odot}}\right)^{-1/2}\left(\frac{r_{\perp}}{2.5~{\rm AU}}\right)^{3/2} \; , \label{eqn:keplers_law}
\end{align}
where we have assumed that $m_p << M_l$. The radial velocity semi-amplitude of this typical microlensing planet is thus
\begin{align}
    K_{\rm typ} \approx & {} \; \left(\frac{2\pi G}{P_{\rm typ}}\right)^{1/3} \frac{m_p\sin{i}}{M_l^{2/3}} \nonumber \\
    \approx & {} \; 5~{\rm m~s^{-1}}\left(\frac{M_l}{\rm 0.5~M_{\odot}}\right)^{1/2}\left(\frac{q}{5\times 10^{-4}}\right) \nonumber \\
    & {} \; \times \left(\frac{r_{\perp}}{2.5~{\rm AU}}\right)^{-1/2} \; , \label{eqn:vel_semi_amp}
\end{align}
for a planet of mass $m_p \sim 0.26~M_{\rm Jup}$ on a circular orbit at the median $\sin{i} \approx 0.866$.

Radial velocity surveys of low-mass stars (particularly M dwarfs) have reached time baselines comparable to this $7$~yr period, and sensitivities of a few ${\rm m~s^{-1}}$ are routinely achieved, so we can compute the SNR, via equation (\ref{eqn:snr_small_p}), to which such surveys should be able to detect these sorts of planets. We examine a fiducial RV survey which consists of a sample of stars each with roughly $N\approx 30$ epochs and uncertainties $\sigma\approx 4~{\rm m~s^{-1}}$ (which consists of, for example, $2~{\rm m~s^{-1}}$ precision errors and $3.5~{\rm m~s^{-1}}$ stellar jitter). Equation (\ref{eqn:snr_small_p}) then tells us that our fiducial RV survey should detect planets around low-mass stars with periods of $\sim 7$~yr and $K\sim 5~{\rm m~s^{-1}}$ with $\mathcal{Q} \sim 5$. This would only constitute a marginal detection, but planets with larger mass ratios and/or smaller projected separations would be more readily detectable and should show up in the RV sample. In addition, RV surveys should be beginning to discover trends (i.e. long-term RV variations) in their samples, resulting from the long-period companions microlensing finds. To the extent that the parameters of our fiducial RV survey are reasonable, this implies that microlensing and RV surveys are beginning to overlap a common region of parameter space, warranting a careful, thorough comparison between these two samples.

\section{Distributions of RV Observables for the Typical Microlensing Planet}
\label{sec:typical_dists}
The ultimate goal of any exoplanet demographics survey is to obtain the distribution function $d^n{\rm N_{pl}}/d\{\alpha\}$, where $\left\{\alpha\right\}$ is the set of all $n$ intrinsic, physical parameters on which planet frequency fundamentally depends. These parameters include, but are not necessarily limited to, host star mass, stellar metallicity, distance to host star, planet mass, semi-major axis, and eccentricity: $\left\{M_l, \left[{\rm Fe/H}\right], D_l, m_p, a, e\right\}$. The total number of planets covered by the domain $\left\{\alpha\right\}$, obtained by marginalizing this distribution function over all parameters, is given by
\begin{equation}
    {\rm N_{pl}} = \int_{\alpha_0}d\alpha_0\int_{\alpha_1}d\alpha_1\dotsi\int_{\alpha_n}d\alpha_n\frac{d^n{\rm N_{pl}}}{d\left\{\alpha\right\}}\; .
\end{equation}
This distribution function is the most general, fundamental quantity related to the properties and demographics of exoplanets. If we were able to detect every planet and measure these properties, we could in principal derive this distribution function. However, the number of planets we can actually detect depends on many observable parameters affecting detectability, some of which belong to the set $\left\{\alpha\right\}$ or are functions of the parameters in the set $\left\{\alpha\right\}$. We group these observable parameters affecting detectability into the set $\left\{\beta\right\}$, containing $k$ elements including (for RV surveys), but not necessarily limited to, velocity semi-amplitude, orbital period, stellar radius, inclination, mean anomaly, and argument of periastron: $\left\{K, P, R_{\star}, i, M_0, \omega\right\}$.

In reality, the number of planets actually found in any exoplanet survey, be it RV, microlensing or other, is 
\begin{align}
    {\rm N_{pl, obs}} = & {} \displaystyle \int_{\beta_0}d\beta_0\int_{\beta_1}d\beta_1\dotsi\int_{\beta_k}d\beta_k\frac{d^k{\rm N_{pl}}}{d\left\{\beta\right\}} \nonumber \\
    & {} \displaystyle \times \prod_{j=0}^k\Phi\left(\beta_j\right)\; , \label{eqn:n_pl_obs}
\end{align}
where $\Phi\left(\left\{\beta\right\}\right)$ is the set of ``efficiency'' functions for each parameter in $\left\{\beta\right\}$ that a particular survey suffers. The distribution function $\frac{d^k{\rm N_{pl}}}{d\left\{\beta\right\}}$ describes the properties of planets in terms of detectability parameters and is given by
\begin{align}
    \frac{d^k{\rm N_{pl}}}{d\left\{\beta\right\}} = & {} \displaystyle \int_{\alpha_0}d\alpha_0\int_{\alpha_1}d\alpha_1\dotsi\int_{\alpha_n}d\alpha_n\frac{d^n{\rm N_{pl}}}{d\left\{\alpha\right\}} \nonumber \\
    & {} \displaystyle \times \prod_{j=0}^k\delta\left(\beta_j\left(\left\{\alpha\right\}\right)-\beta'\right)\; .
\end{align}

Thus, even if we knew the exact form of our ``efficiency'' functions, $\Phi\left(\left\{\beta\right\}\right)$, we are still unable to derive the true distribution function of planets in terms of \emph{only} the intrinsic parameters affecting planet frequency, $\left\{\alpha\right\}$, because any given survey is not sensitive to the full parameter space spanned by all observable parameters $\left\{\beta\right\}$. Indeed, the parameters we can partially marginalize out depends on the type of survey, and presents a great challenge when trying to synthesize exoplanet demographics from multiple different surveys. Regardless, comparing and synthesizing data from multiple exoplanet detection data methods is 
the \emph{only} way to cover a maximal amount of observable space and thus get as close to we can to obtaining the true distribution function, $d^n{\rm N_{pl}}/d\{\alpha\}$, ultimately providing key empirical constraints necessary to learn about planet formation. This study is a step in that direction, as we aim to describe a comparison between microlensing and RV surveys.

\subsection{Circular Orbits}
\label{subsec:circular_orbits}
In \S~\ref{sec:OoM}, we examined the ``typical'' microlensing detection and assumed fixed values of $i$, $\theta$, and thus $a$, thus deriving one $\left(K,P\right)$ pair. In this section, we perform the next step of deriving the distributions of $K$ and $P$ resulting from marginalizing over the unknown orbital parameters, fixing the eccentricity to zero, $e=0$. We are still assuming fixed $M_l=0.5~M_{\odot}$, $q=5\times10^{-4}$, $s=1$, and $r_{\perp}=2.5~$AU, i.e. our ``typical'' microlensing-detected planetary system. For a fixed source distance of $D_s=8~$kpc, $r_{\perp}=2.5~$AU and $M_l=0.5~M_{\odot}$ implies $D_l\sim 4~$kpc. This yields a distribution of $K$ and $P$ formally given by
\begin{align}
    & {} \frac{d^6N_{\rm pl}}{dKdPdM_ldD_ld\log{q}d\log{s}} = \displaystyle \int_{\left\{\alpha\right\}}d\left\{\alpha\right\} \nonumber \\
    & {} \hspace{0.5in} \times \frac{d^n{\rm N_{pl}}}{d\left\{\alpha\right\}}\delta\left(K\left(m_p, i, M_l, a\right)-K'\right) \nonumber \\
    & {} \hspace{0.5in} \times \delta\left(P\left(M_l, m_p, a\right)-P'\right)\delta\left(M_l-0.5~{\rm M_{\odot}}\right) \nonumber \\
    & {} \hspace{0.5in} \times \delta\left(D_l-4~{\rm kpc}\right)\delta\left(\log{q}+3.3\right)\delta\left(\log{s}\right)\; ,\label{eqn:circ_dist}
\end{align}
where, for now, we assume
\begin{equation}
    \frac{d^n{\rm N_{pl}}}{d\left\{\alpha\right\}} = \frac{d{\rm N_{pl}}}{di}\frac{d{\rm N_{pl}}}{da}\frac{d{\rm N_{pl}}}{dM_0}\frac{d^2{\rm N_{pl}}}{d\log{q}~d\log{s}}\frac{d{\rm N_{pl}}}{dM_l}\frac{d{\rm N_{pl}}}{dD_l}\; .
\end{equation}

Since we are only concerned with circular orbits at the moment, we have a delta function at an eccentricity of zero and we do not marginalize over the argument of pericenter, $\omega$, as it is undefined for circular orbits. The marginalization over the longitude of the ascending node, $\Omega$, is also unnecessary, as this angle is degenerate in a microlensing detection. This will also be true when we consider eccentric orbits, although in that case, we will need to marginalize over a range of eccentricities and arguments of pericenter.

Rather than attempting to derive an analytic form of equation (\ref{eqn:circ_dist}), we determine the posterior distribution numerically. This is accomplished by creating an ensemble of circular orbits, each described by a set of parameters $\left\{a, i, M_0\right\}$, drawn from mostly trivial priors. For each orbit, the mean anomaly, $M_0$, is drawn from a uniform distribution in the range $\left[0,2\pi\right)$ and $\cos{i}$ is drawn uniformly from the range $\left[0,1\right]$. We adopt a distribution in the form of \"{O}pik's law \citep{1924PTarO..25f...1O}, i.e. log-uniform, for the semimajor axis, consistent with the observed intrinsic distribution of planet separations from RV studies \citep{2008PASP..120..531C}. We considered other power-law forms for $dN_{\rm pl}/da$, but found that our results did not significantly change for any reasonable exponents.

We compute the projected separation for each orbit as $\tilde{r} = \left(x^2+y^2\right)^{1/2}$, where $\left(x,y\right)$ are the coordinates of the planet in the plane of the sky, with the host star at the origin. These coordinates are calculated from
\begin{align}
    x = & {} AX +FY \label{eqn:proj_sep_x} \\
    y = & {} BX + GY \; ,\label{eqn:proj_sep_y}
\end{align}
where $\left(X,Y\right)$ are the elliptical rectangular coordinates in the plane of the orbit,
\begin{align}
    X = & {} \cos{E} - e \label{eqn:proj_sep_bigX} \\
    Y = & {} \left(1-e^2\right)^{1/2}\sin{E} \; ,\label{eqn:proj_sep_bigY}
\end{align}
and where $A$, $B$, $G$, and $F$ are the Thiele-Innes constants \citep{1883AN....104..245T,1960QB821.B56......,1978GAM....15.....H} given by
\begin{align}
    A = & {} a\left(\cos{\omega}\cos{\Omega} - \sin{\omega}\sin{\Omega}\cos{i}\right) \label{eqn:proj_sep_A} \\
    B = & {} a\left(\cos{\omega}\sin{\Omega} + \sin{\omega}\cos{\Omega}\cos{i}\right) \label{eqn:proj_sep_B} \\
    F = & {} -a\left(\sin{\omega}\cos{\Omega} + \cos{\omega}\sin{\Omega}\cos{i}\right) \label{eqn:proj_sep_F} \\
    G = & {} a\left(\cos{\omega}\cos{\Omega}\cos{i} - \sin{\omega}\sin{\Omega}\right) \; .\label{eqn:proj_sep_G}
\end{align}
The eccentric anomaly, $E$, is calculated from Kepler's equation, $M_0 = E-e\sin{E}$. In the case of circular orbits, the solution is trivial. There are two other constants necessary to completely describe the orbit in rectangular coordinates, $C = a\sin{\omega}\sin{i}$ and $H = a\cos{\omega}\sin{i}$. These are not necessary for computing the projected separation, and we include them only for the sake of completeness \citep[for more information, see pg. 68 of ][]{2011exha.book.....P}. Equations (\ref{eqn:proj_sep_A})--(\ref{eqn:proj_sep_G}) are general equations, but in our case, they greatly reduce in form because the projected separation, $\tilde{r}$, is independent of the longitude of the ascending node\footnote{This is also why this angle is degenerate in a microlensing detection; microlensing is sensitive to a given projected separation.}, $\Omega$. Also, for now, the eccentricity is set to zero. In the next section, this will no longer be the case.

\begin{figure*}
\epsscale{0.9}
\plotone{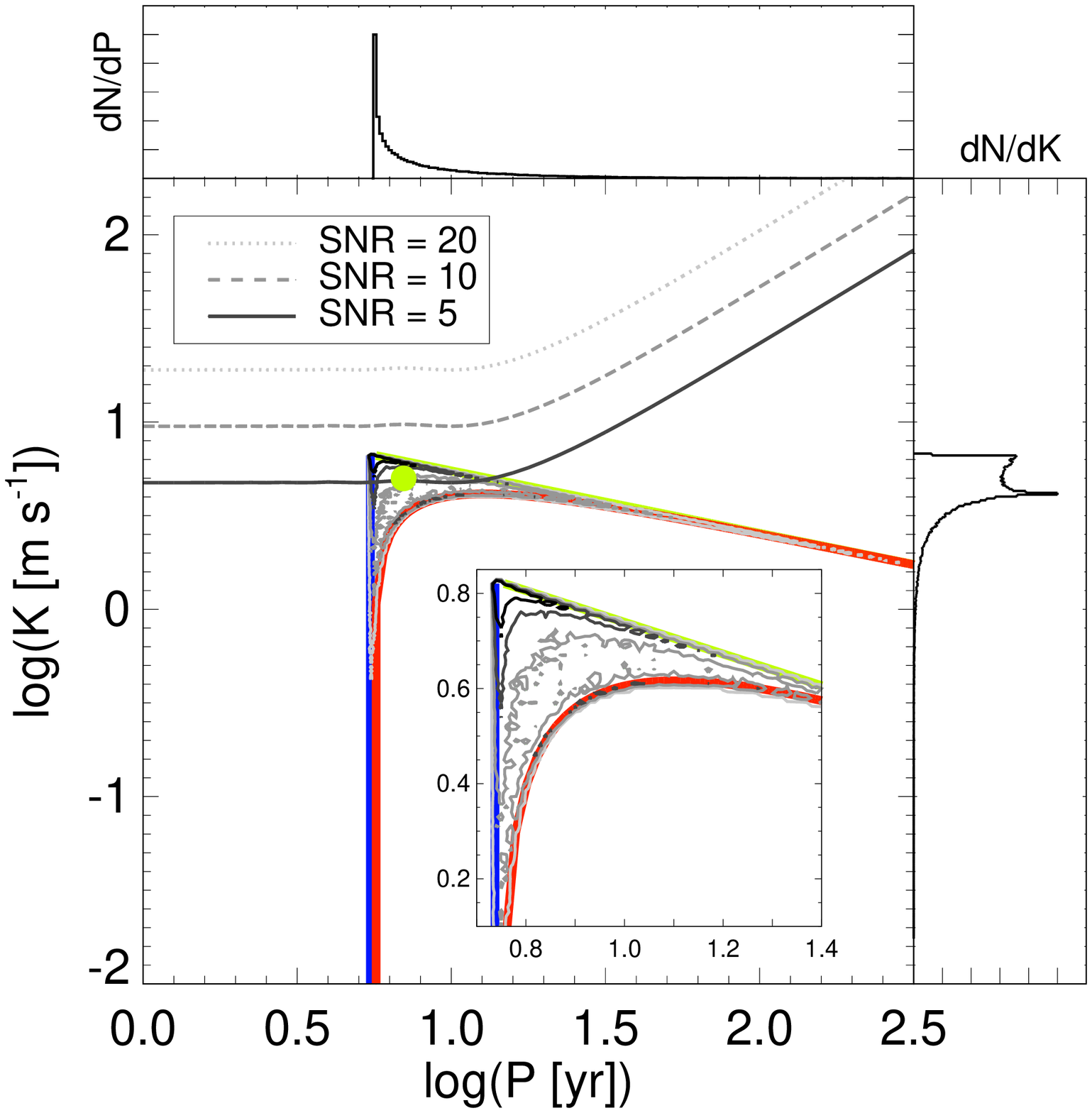}
\caption{Contours of the probability density of $K$ and $P$ for the typical microlensing planet detection ($M_l\sim0.5~M_{\odot}$, $m_p\approx 0.26~M_{\rm Jup}$, $r_{\perp}\sim 2.5~{\rm AU}$) on circular orbits are shown in greyscale. The contour levels, going from grey to black, are $1\%$, $5\%$, and $10\%$ of the peak density. The filled yellow circle represents where the typical microlensing planet lies in this parameter space at the median inclination and mean anomaly ($K_{\rm typ}\sim 5~{\rm m~s^{-1}}$, $P_{\rm typ}\sim 7~$yr). The inset in the lower right is a zoom-in of the central region. The greyscale curves represent contours of constant $\mathcal{Q}$ (i.e. SNR) given by equation (\ref{eqn:snr_full}), assuming the parameters of our fiducial RV survey ($N=30$, $T=10~$yr and $\sigma=4~{\rm m~s^{-1}}$). The colored lines represent analytic constraints on the allowed parameter space of the typical microlensing planet on a circular orbit. The vertical blue line represents the minimum allowed period, given by equation (\ref{eqn:p_low_lim}). The upper bound on $K$, drawn in yellow, is described by equation (\ref{eqn:k_up_lim}), and the lower bound on $K$, drawn in red, is described by equation (\ref{eqn:k_low_lim}). The top panel is the $P$ distribution marginalized over $K$, whereas the right panel is the distribution of $K$ marginalized over $P$.
  \label{fig:kp_contours_typical}}
\end{figure*}

Once we compute the projected separations, we ``pick out'' those orbits with $\tilde{r}$ consistent with that of the microlensing detection, $r_{\perp}$. Specifically, the criterion we employ is $\left|\tilde{r} - r_{\perp}\right| < 0.001R_E$, and is a numerical representation of the Dirac delta function in $r_{\perp}$ of equation (\ref{eqn:circ_dist}). We then compute the joint distribution of $K$ and $P$ from those orbits that survive. This joint distribution is shown in figure \ref{fig:kp_contours_typical}, along with constant SNR contours (equation \ref{eqn:snr_full}) assuming $N=30$, $T=10~$yr and $\sigma=4~{\rm m~s^{-1}}$.

This figure illustrates an important point. In \S~\ref{sec:OoM} we found that the typical microlensing planet (at the median inclination and mean anomaly, and thus at fixed semimajor axis) should be detectable by our fiducial RV survey at a SNR of about 5. However, when the orbital parameters $\left\{a, i, M_0\right\}$ are allowed to vary, we find that the typical microlensing planet is not always detectable by our fiducial RV survey. The spread in both $K$ and $P$ results completely from variance in the orbital parameters $\left\{a, i, M_0\right\}$ and thus only a subset of realizations of the typical microlensing planet turn out to be detectable.

Figure \ref{fig:kp_contours_typical} also demonstrates how the sensitivities of both the microlensing and RV methods are imprinted on the distribution of RV observables we derive. There are two competing effects that produce a double-peaked distribution in $K$ for a given $P$. This is reflected in the structure of the inner contour lines. These two effects arise from a preference for edge-on and face-on orbits over intermediate inclinations. The first effect is that edge-on orbits are preferred since the distribution of $\cos{i}$ is uniform on $\left[0,1\right]$ for a randomly-oriented, circular orbit. This tends to produce velocity semi-amplitudes near the maximum possible value and causes the over-density in parameter space bounded by the inner, darker contour at larger $K$ values in figure \ref{fig:kp_contours_typical}. The other effect is that there is more phase space for a microlensing detection near face-on orbits since microlensing is sensitive to projected separations near the Einstein radius, producing velocity semi-amplitudes near the minimum possible value. Planets with face-on orbits spend more time near the Einstein ring than those with edge-on orbits and are thus more likely to produce a detectable perturbation during a microlensing event. The over-density in parameter space due to this effect is bounded by the inner, lighter shaded contour at relatively lower $K$ values in figure \ref{fig:kp_contours_typical}. Thus, after marginalizing $d^2N_{\rm pl}/(dKdP)$ over the period, the independent distribution of $K$ is imprinted with a feature, a peak, in the interior of its allowed range, resulting from the superposition of these two effects (shown in the right panel of figure \ref{fig:kp_contours_typical}).

There are distinct physical reasons for the shapes of these contour lines resulting from constraints on the allowed parameter space. Here we derive these constraints for the general case, including eccentric orbits, as we will need to return to the following relations when including eccentricity. The minimum period that a microlensing planet can have is realized when the projected separation is equal to the apastron of the planetary orbit, $r_{\perp} = a\left(1+e\right)$. From Kepler's 3rd law, this period is given by
\begin{equation}
    P_{\rm min} = 2\pi\left(\frac{r_{\perp}^3}{GM_l}\right)^{1/2}\left(1+e\right)^{-3/2} \; . \label{eqn:p_low_lim}
\end{equation}

The upper bound on $K$ for a given period results from the case that produces the largest radial velocities: an edge-on orbit ($i = 90^{\circ}$). This upper bound on $K$ is simply
\begin{equation}
    K_{\rm max} = \left(\frac{2\pi G}{P}\right)^{1/3}\frac{m_p}{M_l^{2/3}}\left(1-e^2\right)^{-1/2} \; . \label{eqn:k_up_lim}
\end{equation}
This minimum bound on $K$ for a given $P$ corresponds to the minimum inclination allowed by the projected separation. The inclination is minimized when the projected separation is equal to the periastron of the planetary orbit, $r_{\perp} = a\left(1-e\right)$, and the inclination is given implicitly by
\begin{equation}
    \cos{i} = \frac{r_{\perp}}{a\left(1-e\right)} \; ,
\end{equation}
which we can solve explicitly for $i$ and rewrite in terms of the period, $P$:
\begin{equation}
    i = \cos^{-1}\left(\frac{r_{\perp}}{\left(1-e\right)}\left[\frac{GM_l}{4\pi^2}P^2\right]^{-1/3}\right) \; .
\end{equation}
The lower bound on $K$ for a given period and measured projected separation is then
\begin{align}
    K_{\rm min} = & {} \left(\frac{2\pi G}{P}\right)^{1/3} \frac{m_p}{M_l^{2/3}} \nonumber \\
    & {} \times \left\{1 - \left[\frac{r_{\perp}}{\left(1-e\right)}\right]^2\left[\frac{GM_l}{4\pi^2}P^2\right]^{-2/3}\right\}^{1/2}\; . \label{eqn:k_low_lim}
\end{align}
The maximum of the curve defined by equation (\ref{eqn:k_low_lim}) occurs at
\begin{equation}
    \left.P\right|_{K_{\rm min}={\rm max}\left(K_{\rm min}\right)} = \frac{2\pi3^{3/4}}{\left(GM_l\right)^{1/2}}\left(\frac{r_{\perp}}{1-e}\right)^{3/2} \label{eqn:p_k_max_k_min}
\end{equation}
and has the value
\begin{equation}
   {\rm max}\left(K_{\rm min}\right) = \frac{2^{1/2}m_p}{3^{3/4}}\left(\frac{1-e}{r_{\perp}}\right)^{1/2}\left(\frac{G}{M_l}\right)^{1/2} \; , \label{eqn:max_k_min_peak}
\end{equation}
which corresponds to the location of the peak in $dN_{\rm pl}/dK$ shown in figure \ref{fig:kp_contours_typical}. Note that equation (\ref{eqn:k_low_lim}) contains within it a requirement that
\begin{equation}
    P \geq 2\pi\left(\frac{r_{\perp}^3}{GM_l}\right)^{1/2}\left(1-e\right)^{-3/2} \label{eqn:min_p_faceon}
\end{equation}
for $K$ to be real. When equation (\ref{eqn:min_p_faceon}) is an equality, this corresponds to the period for a perfectly face-on orbit, where $K$ is exactly zero. For circular orbits, this period is equivalent to the minimum period for any microlensing planet, given by equation (\ref{eqn:p_low_lim}).

We plot the analytical constraints defined by equations (\ref{eqn:p_low_lim}), (\ref{eqn:k_up_lim}), and (\ref{eqn:k_low_lim}) in figure \ref{fig:kp_contours_typical}. For a given microlensing detected planet with measured $r_{\perp}$, $M_l$, and $m_p$, the RV observables $K$ and $P$ it could produce must lie within these bounds. As we will see in the next subsection, the area of this allowed parameter space depends strongly on eccentricity and is a minimum in the case of circular orbits.

\subsection{Eccentric Orbits}
\label{subsec:eccentric_orbits}
We now examine how adding eccentricity changes the form of the posterior joint $\left(K,P\right)$ distribution for the typical microlensing planet defined in \S~\ref{sec:OoM}. The overall process is the same as the circular case, except now we must marginalize over a range of eccentricities and arguments of pericenter. Equation (\ref{eqn:circ_dist}) still holds, but now we assume
\begin{align}
    \frac{d^n{\rm N_{pl}}}{d\left\{\alpha\right\}} = & {} \frac{d{\rm N_{pl}}}{di}\frac{d{\rm N_{pl}}}{da}\frac{d{\rm N_{pl}}}{dM_0}\frac{d^2{\rm N_{pl}}}{d\log{q}~d\log{s}}\nonumber \\
    & {} \times \frac{d{\rm N_{pl}}}{dM_l}\frac{d{\rm N_{pl}}}{dD_l}\frac{d{\rm N_{pl}}}{d\omega}\frac{d{\rm N_{pl}}}{de}\; . \label{eqn:ecc_alpha_terms}
\end{align}

\begin{figure*}
\epsscale{0.9}
\plotone{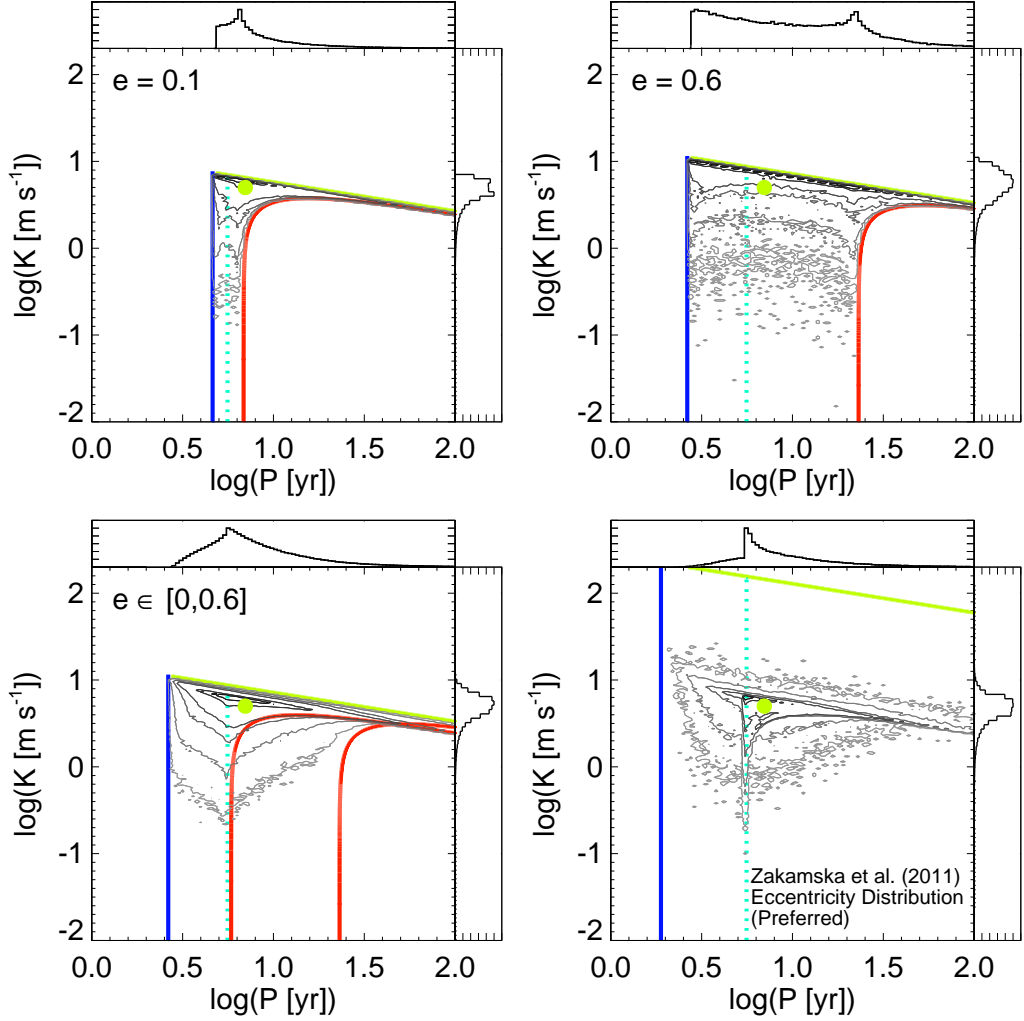}
\caption{Contours of the probability density of $K$ and $P$ for the typical microlensing planet detection ($M_l\sim0.5~M_{\odot}$, $m_p\approx 0.26~M_{\rm Jup}$, $r_{\perp}\sim 2.5~{\rm AU}$) are shown in greyscale. The contour levels, going from grey to black, are $0.1\%$, $1\%$, $5\%$, $10\%$, $30\%$ and $50\%$ of the peak density. The filled yellow circle represents where the typical microlensing planet lies in this parameter space at the median inclination and mean anomaly and on a circular orbit ($K_{\rm typ}\sim 5~{\rm m~s^{-1}}$, $P_{\rm typ}\sim 7~$yr). In each panel, the colored lines represent the analytical constraints on the allowed parameter space. The vertical blue lines represent the minimum allowed period, given by equation (\ref{eqn:p_low_lim}). The upper bounds on $K$, drawn in yellow, are described by equation (\ref{eqn:k_up_lim}), and the lower bounds on $K$, drawn in red, are described by equation (\ref{eqn:k_low_lim}). The vertical cyan line represents the minimum allowed period for a circular orbit. These plots illustrate how eccentricity greatly affects the area of the allowed parameter space in comparison to the circular case (see figure \ref{fig:kp_contours_typical}). Each panel uses a different prior on the eccentricity. The uniform prior used in the lower left panel allows for both circular and eccentric orbits, so the allowed parameter space is bounded by components of two minimum $K$ constraints, one from $e=0$ and $e=0.6$.
  \label{fig:kp_contours_ecc_test}}
\end{figure*}

As in the circular case, we choose a prior in $\cos{i}$ that is uniform on $\left[0,1\right]$ and a log-uniform distribution for semimajor axis. The priors of the mean anomaly, $M_0$, and the argument of pericenter, $\omega$, are uniform on $\left[0,2\pi\right)$. At periastron, we have $M_0=0$, while $M_0=\pi$ corresponds to apastron. We examine four different priors on eccentricity to see how sensitive the $K$ and $P$ distributions are to the eccentricity distribution. Figure \ref{fig:kp_contours_ecc_test} shows the distributions using the various eccentricity priors we explore. First, we try a couple priors that are delta functions at given values of eccentricity (i.e. all planets have a specific value of eccentricity) to explore the effects orbits of fixed eccentricity will have on the resultant $K$ and $P$ distributions. The eccentricity prior in the top left panel of figure \ref{fig:kp_contours_ecc_test} is $dN_{\rm pl}/de \equiv \delta(e-0.1)$ and that of the top right panel is $dN_{\rm pl}/de \equiv \delta(e-0.6)$. It is clear from these two panels alone that eccentricity can affect the joint distribution of $K$ and $P$, as the allowed parameter space can be increased significantly relative to the circular case. The median values of this distribution, $\left(K_{\rm med},P_{\rm med}\right)$, are $\left(4.53~{\rm m~s^{-1}}, 7.08~{\rm yr}\right)$ and $\left(5.32~{\rm m~s^{-1}}, 9.24~{\rm yr}\right)$, for the top left ($e=0.1$) and right ($e=0.6$) panels, respectively. For comparison, the median values in the case of only circular orbits, where $e \equiv 0$, are $\left(4.50~{\rm m~s^{-1}}, 6.94~{\rm yr}\right)$, as seen in figure \ref{fig:kp_contours_typical}. It is clear that the median and spread of the $P$ distribution are more sensitive to eccentricity than those of the $K$ distribution. We explain this below.

Next, we try an eccentricity prior uniform in the range $\left[0,0.6\right]$, producing the median values $\left(4.75~{\rm m~s^{-1}}, 7.86~{\rm yr}\right)$, as shown in the bottom left panel. When the eccentricity is allowed to vary, the constraints on the allowed parameter space are not as simple as before. The lower bound on $K$ in equation (\ref{eqn:k_low_lim}) varies between the value for circular orbits, $e=0$, and that for the maximum eccentricity, $e=e_{\rm max}=0.6$.

Finally, we also consider the intrinsic (de-biased) eccentricity distribution determined by \citet{2011MNRAS.410.1895Z} from the RV detection sample of \citet{2006ApJ...646..505B}. The cumulative distribution function of this distribution is shown in figure 12 of \citet{2011MNRAS.410.1895Z}, representing the sum of two planet populations: one on circular orbits (38\% of all planets) and the remaining a ``dynamically active'' distribution with eccentricities following the form of \citet{2008ApJ...686..603J}:
\begin{equation}
    \frac{dN_{\rm pl}}{de} \propto e \exp{\left[-\frac{1}{2}\left(\frac{e}{0.3}\right)^2\right]} \; . \label{eqn:jt_ecc_dist}
\end{equation}
When using this prior we find a sharp feature in the distribution of $P$, corresponding to circular orbits. This pile-up corresponds to a similar feature in the semimajor axis distribution, resulting from the fact that the \citet{2011MNRAS.410.1895Z} eccentricity distribution has a pile-up of planets on circular orbits relative to planets on eccentric orbits. When using this eccentricity prior, we find the median values of $\left(4.71~{\rm m~s^{-1}}, 7.51~{\rm yr}\right)$ for $K$ and $P$, respectively.

In summary, including a range of eccentricities does not significantly change the median values of $K$ or $P$, but does serve to significantly broaden both distributions, particularly that of $P$. Microlensing is most sensitive to planetary perturbations from planets near the Einstein radius, and so is most sensitive to face-on orbits. For finite eccentricities, projected separations from $a(1-e)$ to $a(1+e)$ can satisfy the requirement that $r_{\perp}/R_E=a(1\pm e)/R_E=s$, rather than just $a$. The distribution of $P$ is thus sensitive to variations in eccentricity, whereas at fixed $s$, $q$, $M_l$, $i$, the velocity semi-amplitude scales as $K\propto P^{-1/3}\left(1-e^2\right)^{-1/2}$, such that changes in $e$ have two effects that work in opposite directions on $K$ (and so approximately cancel), meaning that the distribution of $K$ is much less sensitive to $e$. However, we will see in future sections that when we marginalize over a range of $s$, $q$, $M_l$, and $i$, sharp features in the distribution of $P$ are diminished and that variations in these parameters tend to spread out both $K$ and $P$.

The allowed RV parameter space into which a planet detected by microlensing can be mapped is severely limited in the case of circular orbits, but allowing for eccentric orbits relaxes these constraints. When eccentricities are allowed to be non-zero, a much larger region of $\left(K, P\right)$ space is made available, as illustrated in figure \ref{fig:kp_contours_ecc_test}. The colored lines show these relaxed constraints, influenced mainly by the upper bound on the eccentricity prior. When the eccentricity is allowed to vary, the peak of the $P$ distribution occurs at the minimum possible period for a circular orbit from equation (\ref{eqn:p_low_lim}). As the eccentricity is allowed to get arbitrarily close to 1, the entire parameter space becomes available. However, we place a conservative upper limit on the eccentricity to be $0.999$.

We henceforth adopt as our eccentricity prior that of \citet{2011MNRAS.410.1895Z}. Since this eccentricity distribution is determined from RV planet detections around FGK dwarfs, many of which must have migrated, it is unclear if the M dwarf planet population follows this distribution. However, unless a very large fraction of large-separation planets have very high eccentricities, our results will not depend sensitively on the precise eccentricity prior we adopt.

\section{Microlensing Parameter Distributions}
\label{sec:marg}
Previously, we have considered the joint distribution of $K$ and $P$ for the typical microlensing planet detection, i.e. for fixed values of $M_l$, $D_s$, $D_l$ (and so $R_E$), $q$, and $s$ (and so $m_p$ and $r_{\perp}$). In reality, the microlensing detections have a distribution of host star properties\footnote{There is also some distribution of source distances, $D_s$, however since a vast majority of sources are located in the bulge, the range of $D_s$ is smaller than that of $D_l$. Furthermore, because a majority of lenses are also in the bulge and because the lens must be in front of the source, the range of $D_s$ is even further constrained, somewhere behind the Galactic center and in front of the back edge of the bulge (a range of a few kpc).}, $M_l$ and $D_l$, as well as a distribution of planet properties, $q$ and $s$. Considering the full range of microlensing systems by varying $M_l$, $D_l$, $q$, and $s$, in addition to the orbital parameters, yields the distribution given formally as
\begin{align}
    & {} \frac{d^6N_{\rm pl}}{dKdPdM_ldD_ld\log{q}d\log{s}} = \displaystyle \int_{\left\{\alpha\right\}}d\left\{\alpha\right\} \nonumber \\
    & {} \hspace{0.35in} \times \frac{d^n{\rm N_{pl}}}{d\left\{\alpha\right\}}\delta\left(K\left(m_p, i, M_l, a\right)-K'\right) \nonumber \\
    & {} \hspace{0.35in} \times \delta\left(P\left(M_l, m_p, a\right)-P'\right)\delta\left(M_l-M_l'\right)\delta\left(D_l-D_l'\right) \nonumber \\
    & {} \hspace{0.35in} \times \delta\left(\log{q}-\log{q'}\right)\delta\left(\log{s}-\log{s'}\right)\; ,\label{eqn:full_mlens_dist}
\end{align}
where $d^n{\rm N_{pl}}/d\{\alpha\}$ is given by equation (\ref{eqn:ecc_alpha_terms}). We note that
\begin{equation}
    \frac{dN_{\rm pl}}{dM_l}\frac{dN_{\rm pl}}{dD_l} \propto \displaystyle \int \int \frac{d^4d\Gamma}{dD_ldM_ld^2\boldsymbol{\mu}}\Phi\left(t_E\right)d^2\boldsymbol{\mu}\; ,
\end{equation}
where $d^4d\Gamma/(dD_ldM_ld^2\boldsymbol{\mu})$ is the event rate of a given microlensing event, $\Phi\left(t_E\right)$ is an event timescale, $t_E$, selection function, and $\boldsymbol{\mu}$ is the lens-source relative proper motion. We describe the event rate and all associated parameters in the following sections.

Only a subset of microlensing events have measurements of (or even constraints on) the lens mass and distance and thus we do not know the true distribution of $M_l$ and $D_l$ for the microlensing sample. We choose to ignore the subset of such measurements that have been made and instead adopt reasonable assumptions for these distributions based on a Galactic model, taking care to weight each microlensing event by its expected event rate. We then assume distributions of $q$ and $s$ determined by microlensing surveys. Furthermore, we must also weight by an efficiency function on the event timescale $t_E$ of each microlensing event because we will later be normalizing our answers to the occurrence rate found by \citet{2010ApJ...720.1073G}. This efficiency function is meant to reproduce the slight bias towards longer timescale events in the \citet{2010ApJ...720.1073G} sample.

In this section, we describe how to generate a planetary microlensing event characterized by the parameters $M_l$, $D_l$, $q$, and $s$ along a given sight line, $\left(l,b\right)$, occurring at a given time of the year and with an appropriate event rate. We then determine the distribution of $K$ and $P$, as before. We repeat this process varying all the different parameters and stack the resultant distributions of $K$ and $P$, weighting each event by its event rate, $d^4d\Gamma/(dD_ldM_ld^2\boldsymbol{\mu})$, and a timescale, $t_E$, bias. This yields a marginalized distribution that represents a full, weighted mapping of microlensing planet parameter space into radial velocity planet parameter space, i.e. our best estimate of the expected probability distribution of the radial velocity observables for a sample of stars hosting analogs to the planets inferred from microlensing surveys.

\subsection{Microlensing Event Rate}
\label{subsec:event_rate}
The rate equation for a microlensing event with a given $D_l$, $M_l$, and $\boldsymbol{\mu}$ toward a given location $\left(l, b\right)$ is
\begin{equation}
    \frac{d^4\Gamma}{dD_ldM_ld^2\boldsymbol{\mu}} = 2R_Ev_{\rm rel}\nu\frac{d\Gamma}{d\boldsymbol{\mu}}\frac{d\Gamma}{dM_l} \; , \label{eqn:mlens_event_rate}
\end{equation}
where $\nu$ is the local number density of lenses, $R_E$ is the physical Einstein radius, $v_{\rm rel}$ is the lens-source relative velocity, $d\Gamma/d\boldsymbol{\mu}$ is the two-dimensional probability density for a given lens-source relative proper motion, $\boldsymbol{\mu}$, and $d\Gamma/dM_l$ is the mass function. We will use this event rate equation as part of a statistical weight for each of our simulated microlensing events. The lens-source relative velocity in physical units is given by
\begin{equation}
    v_{\rm rel}\left(D_l, \left|\boldsymbol{\mu}\right| \right) = \left|\boldsymbol{\mu}\right|D_l\label{eqn:v_rel} \; .
\end{equation}
Each of our simulated events is associated with a line-of-sight in Galactic longitudes and latitudes, $\left(l,b\right)$. Combined with the distance to the lens, $D_l$, the Galactic longitude and latitude yield the precise location of the event in the Galactic frame, $\left(x,y,z\right)$, which is necessary for computing the density of lenses, $\nu\left(x,y,z\right)$, at the location of the lens.

\subsubsection{Density of Lenses}
\label{subsubsec:density_of_lenses}
We need to know the number density of lenses at particular locations in the Galaxy in order to compute event rates. This number density, $\nu\left(x,y,z\right)$, is normalized to some fiducial mass (which is a constant, i.e. same for all events), and thus we have that $\nu\left(x,y,z\right) \propto \rho\left(x,y,z\right)$. Since we are only interested in the relative probabilities of our ensemble of microlensing events, the absolute normalization does not matter, and we can instead compute the mass density in place of the number density in the event rate equation. The mass function, $d\Gamma/dM_l$, in the event rate equation provides the relative weighting for the various individual lens masses. In this section, we describe the mass density models we adopt for the disk and the bulge.

We use a two-component mass density distribution, including a disk and a barred, anisotropic bulge. As in \citet{2011A&A...529A.102B}, we assume the disk has cylindrical symmetry with a 1~kpc hole centered at $R_0=8$~kpc. We also ignore events with disk lenses with distances between $9-10~$kpc to eliminate the small fraction of events that arise from disk lenses with apparent retrograde motions due to their location on the far side of the disk (i.e. opposite side of the bulge as the Sun) which thus have very large lens-source relative proper motions and produce short-timescale events that are not observed. Our disk model has a ``double--exponential'' form, given by equations 4 and 6 from \citet{2001ApJ...555..393Z}:
\begin{align}
    & {} \rho_{\rm disk}\left(x,y,z\right) = \rho_{\rm 0, disk}\; \text{exp}\left(-\frac{R-R_0}{H}\right) \nonumber \\
    & {} \hspace{0.42in} \times \left[\left(1-\beta\right)\text{exp}\left(-\frac{\left|z\right|}{h_1}\right) + \beta\text{exp}\left(-\frac{\left|z\right|}{h_2}\right)\right]\; , \label{eqn:disk_model}
\end{align}
where
\begin{equation}
    R = \left(x^2+y^2\right)^{1/2}\; ,
\end{equation}
and $R_0 = 8.0$~kpc is the distance between the Sun and the Galactic Center (GC). The coordinates $\left(x,y,z\right)$ have their origin at the Galactic center with the longest axis, $x$, corresponding to the Sun-GC axis. The $y$-axis is perpendicular to the $x$-axis and lies within the Galactic plane, while $z$-axis is perpendicular to the Galactic plane. The $\left(x,y,z\right)$ coordinates can be computed from the $\left(l,b\right)$ position of a microlensing event via the equations
\begin{align}
    x = & {} R_0 - D_l\cos(l)\cos(b) \label{eqn:x_gal} \\
    y = & {} D_l\sin(l)\cos(b) \label{eqn:y_gal} \\
    z = & {} D_l\sin(b)\; , \label{eqn:z_gal}
\end{align}
The best-fit parameters for this disk model, derived from Hubble Space Telescope star counts and presented in the ``All Data, CMR (2)'' section of Table 3 of \citet{2001ApJ...555..393Z}, are $h_1 = 0.156$~kpc, $h_2 = 0.439$~kpc, $\beta = 0.381$ and $H = 2.75$~kpc.

The bulge model we choose is a ``boxy'' Gaussian, limited to radii of $R\leq 3$~kpc, with a modified radial coordinate that allows for a triaxial morphology. It is model G2, with $R_\text{max} = 5.0$~kpc, from equation 3b of \citet{1995ApJ...445..716D}, and has the form
\begin{equation}
    \rho_{\rm bulge}\left(x',y',z'\right) = \rho_{\rm 0, bulge}\; \text{exp}\left(-0.5r_s^2\right)\; , \label{eqn:bulge_model}
\end{equation}
where
\begin{equation}
    r_s = \left\{\left[\left(\frac{x'}{x_0}\right)^2 + \left(\frac{y'}{y_0}\right)^2\right]^2+\left(\frac{z'}{z_0}\right)^4\right\}^{1/4}\; .
\end{equation}
The coordinates $\left(x', y', z'\right)$ have their origin at the Galactic center with the longest axis, $x'$, rotated 20$^{\circ}$ from the Sun-GC axis toward positive latitude and $z'$ is the shortest axis. These coordinates are the same as the $\left(x,y,z\right)$ coordinates used in the disk model, just rotated by 20$^{\circ}$ about the $z$-axis via the equations
\begin{align}
    x' = & {} \; x\cos(20^{\circ}) + y\sin(20^{\circ}) \label{eqn:x_p} \\
    y' = & {} \; -x\sin(20^{\circ}) + y\cos(20^{\circ}) \label{eqn:y_p} \\
    z' = & {} \; z\; . \label{eqn:z_p}
\end{align}
The best-fit parameters for this bulge model, derived from near-infrared images of the Galactic bulge at 2.2$\mu$m from the Diffuse Infrared Background Experiment (DIRBE) onboard the Cosmic Background Explorer (COBE) satellite and presented in Table 1 of \citet{1995ApJ...445..716D}, are $x_0 = 1.58$~kpc, $y_0 = 0.62$~kpc, and $z_0 = 0.43$~kpc.

We compute the normalizations for these mass density functions as follows. For the disk, we adopt the normalized local stellar column density, $\Sigma_0 = 36~M_{\odot}~{\rm pc}^{-2}$, which includes $28~M_{\odot}~{\rm pc}^{-2}$ in observable stars and white dwarfs \citep{1996ApJ...465..759G,2001ApJ...555..393Z} and another $8~M_{\odot}~{\rm pc}^{-2}$ that is a rough estimate of the column density of brown dwarfs \citep{2003ApJ...592..172H}. Integrating the disk density model perpendicular to the plane of the galaxy at $R=R_0$, and setting this equal to the local stellar column density, we solve for the normalization
\begin{align}
    \rho_{\rm 0, disk} = & {} \Sigma_0\left[2\displaystyle \int_0^{\infty} \left[\left(1-\beta\right)\text{exp}\left(-\left|z\right|/h_1\right)\right.\right. \nonumber \\
    & {} + \Bigl.\left.\beta\text{exp}\left(-\left|z\right|/h_2\right)\right]dz\biggr]^{-1}\; .
\end{align}
This calculation can be carried out analytically and yields $\rho_{\rm 0, disk} = 0.0682~M_{\odot}~{\rm pc}^{-3}$.

\citet{2003ApJ...592..172H} find the total column density of bulge stars towards Baade's Window, including associated brown dwarves and remnants, to be $\Sigma_* = 2086~M_{\odot}~{\rm pc}^{-2}$. We find the normalization for the bulge model by integrating it along the line-of-sight towards Baade's Window and setting this equal to the total column density, yielding
\begin{align}
    \rho_{\rm 0, bulge} = & {} \Sigma_*\left[\displaystyle \int_0^{\infty} \text{exp}\left(-0.5\left\{\left[\left(x'/x_0\right)^2 + \left(y'/y_0\right)^2\right]^2\right.\right.\right. \nonumber \\
    & {} \hspace{0.3in} + \left.\left.\left.\left(z'/z_0\right)^4\right\}^{1/2}\right)d\xi\right]^{-1}\; ,
\end{align}
where $\xi$ is the coordinate along the line-of-sight. The coordinates $x'$, $y'$ and $z'$ are computed from $\left(l=-1^{\circ},b=-3.9^{\circ}\right)$ of Baade's Window and are functions of $\xi$, given by equations (\ref{eqn:x_gal})--(\ref{eqn:z_gal}) and (\ref{eqn:x_p})--(\ref{eqn:z_p}), where $D_l$ is replaced with $\xi$ as we move along the line-of-sight. We numerically compute the integral to determine the normalization factor, $\rho_{\rm 0, bulge} = 1.175~M_{\odot}~{\rm pc}^{-3}$.

We compute the optical depth to a microlensing event towards Baade's window to test our density models. Assuming the source is in the bulge, for disk lenses we find $\tau_{\rm d} \approx 0.66\times 10^{-6}$ and for bulge lenses we find $\tau_{\rm b} \approx 0.93 \times 10^{-6}$. This is in rough agreement with the values of $\left<\tau\right>_0\left({\rm disk}\right) = 0.65\times10^{-6}$ and $\left<\tau\right>_0\left({\rm bulge}\right) = 0.98\times10^{-6}$ reported by \citet{2003ApJ...592..172H}.

\subsubsection{Lens-Source Relative Proper Motion}
\label{subsubsec:lens_source_relative_proper_motion}
Assuming that the velocity distributions of the lenses and sources are Gaussian, we adopt the following form for the two-dimensional probability function for the lens-source relative proper motion (see equation 19 of \citealt{2011A&A...529A.102B} and equation 20 of \citealt{2012ApJ...755..102Y}):
\begin{align}
    \frac{d\Gamma}{d\boldsymbol{\mu}} \propto & {} \frac{1}{\sigma_{\rm \mu,N_{gal}}\sigma_{\rm \mu,E_{gal}}}\text{exp}\left[-\frac{\left(\mu_{\rm N_{gal}} - \mu_{\rm exp,N_{gal}}\right)^2}{2\sigma^2_{\rm \mu,N_{gal}}}\right] \nonumber \\
    & {} \hspace{0.5in} \times \text{exp}\left[-\frac{\left(\mu_{\rm E_{gal}} - \mu_{\rm exp,E_{gal}}\right)^2}{2\sigma^2_{\rm \mu,E_{gal}}}\right]\; , \label{eqn:ls_rel_pm}
\end{align}
where $\boldsymbol{\mu}_{\rm exp}$ is the expected proper motion and the inputs are in Galactic coordinates. More specifically, the coordinates are expressed in the $\left(x,y,z\right)$ system used in section \ref{subsubsec:density_of_lenses}, where the $x$- and $z$-axes point to the Earth and North Galactic pole, respectively. The North Galactic component corresponds to motion along the $z$-axis and the East Galactic component corresponds to motion along the $y$-axis. The expected proper motion, $\boldsymbol{\mu}_{\rm exp}$, takes into account the typical motion of lens and source stars, $\mathbf{v}_{\rm l/s}$, as well as the ``observer's'' motion, which depends on the velocity of the Sun about the Galactic Center, $\mathbf{v}_{\odot}$, and the velocity of the Earth about the Sun, $\mathbf{v}_{\oplus}$. The expected proper motion is thus
\begin{equation}
    \boldsymbol{\mu}_{\rm exp} = \frac{\mathbf{v}_l - \left(\mathbf{v}_{\odot} + \mathbf{v}_{\oplus}\right)}{D_l} - \frac{\mathbf{v}_s - \left(\mathbf{v}_{\odot} + \mathbf{v}_{\oplus}\right)}{D_s}\; .
\end{equation}
The velocity of the Sun in the Galactic frame is $\mathbf{v}_{\odot} = \left(v_{\rm \odot,N_{gal}}, v_{\rm \odot,E_{gal}}\right) = \left(7,12\right)~{\rm km~s^{-1}} + \left(0,v_{\rm rot}\right)$, where $v_{\rm rot} = 220$~km~s$^{-1}$. The line-of-sight projected velocity of the Earth, $\mathbf{v}_{\oplus}$, depends on the time and location of the microlensing event. We first compute the Earth's velocity at a given event time and project it onto the line-of-sight towards the event, yielding a velocity in the Equatorial frame, $\mathbf{v}_{\rm \oplus,eq} = \left(v_{\oplus,N},v_{\oplus,E}\right)$. We then transform this into the Galactic frame, $\mathbf{v}_{\rm \oplus,gal} = \left(v_{\rm \oplus,N_{gal}},v_{\rm \oplus,E_{gal}}\right)$, by the transformation given in equations 22 and 23 of \citet{2011A&A...529A.102B}:
\begin{align}
    v_{\rm \oplus,N_{gal}} = & {} v_{\oplus,N}\cos(59.7^{\circ}) - v_{\oplus,E}\sin(59.7^{\circ}) \\
    v_{\rm \oplus,E_{gal}} = & {} v_{\oplus,N}\sin(59.7^{\circ}) + v_{\oplus,E}\cos(59.7^{\circ})\; .
\end{align}
The typical lens star and source star velocities, $\mathbf{v}_l$ and $\mathbf{v}_s$, depend on where the star is located; either the disk or the bulge. For a lens located in the disk, its velocity is $\mathbf{v}_{\rm l} = \left(v_{\rm l,N_{gal}},v_{\rm l,E_{gal}}\right) = \left(0,v_{\rm rot}\right) - \left(0,10\right)$~km~s$^{-1}$ with a dispersion of $\mathbf{\sigma}_{\rm v} = \left(\sigma_{\rm v,N_{gal}}, \sigma_{\rm v,E_{gal}}\right) = \left(20,30\right)$~km~s$^{-1}$. If the lens is located within the bulge, then its velocity is the same as that of the source, $\mathbf{v}_{\rm l/s} = \left(v_{\rm l/s,N_{gal}},v_{\rm l/s,E_{gal}}\right) = \left(0,0\right)$~km~s$^{-1}$ with a dispersion of $\mathbf{\sigma}_{\rm v} = \left(\sigma_{\rm v,N_{gal}}, \sigma_{\rm v,E_{gal}}\right) = \left(100,100\right)$~km~s$^{-1}$. These velocities and their dispersions are converted to proper motions by dividing by $D_l$ and $D_s$, respectively.

\subsubsection{Mass Function}
\label{subsubsec:mass_function}
As discussed in \S~\ref{sec:mlens_rv_techniques}, in order to obtain a measurement of the lens mass, a measurement of $\theta_E$ (from finite source effects or direct measurement of $\mu_{\rm rel}$) and $\tilde{r}_E$ (from microlens parallax) are required. These measurements are not routine, making it difficult to measure the mass function of lenses. Therefore, as in \citet{2011A&A...529A.102B} and \citet{2012ApJ...755..102Y}, we simply adopt a mass function that is uniform in $\log{M_l}$, i.e. $d\Gamma/dM_l \propto M_l^{-1}$. We set the lower mass limit to be the minimum mass for hydrogen burning, taken to be about $\sim 0.07~M_{\odot}$ \citep{2000ApJ...542..464C}. Nevertheless, there is a subset of the \citet{2010ApJ...720.1073G} events for which there are mass measurements (see table \ref{tab:gould_events}). We perform a two-sample Kolmogorov-Smirnov (K-S) test between the event rate and timescale weighted distribution of $M_l$ for our simulated sample with the distribution of measured $M_l$ for the \citet{2010ApJ...720.1073G} sample and find a $D$-statistic of 0.26 with probability $P\left(D\right)=0.44$. We conclude that our choice of prior on the lens mass distribution produces a consistent distribution of lens masses with the observed sample.

\subsection{Timescale Selection Function}
\label{subsec:timescale_selection_function}
\citet{2011Natur.473..349S} showed that the observed distribution of event timescales from the MOA microlensing survey peaks around $\sim 30~$days and dropping rapidly below $\sim 10~$days and above $\sim 100~$days. In the \citet{2010ApJ...720.1073G} sample of events that were followed up to search for planets, they found they were somewhat biased towards longer timescale events than the underlying \citet{2011Natur.473..349S} sample to a modest statistical significance (see their \S~6.6 for lengthy discussion). The timescale for a microlensing event is given by the Einstein crossing time, or the time it takes the source to cross the Einstein ring of the lens,
\begin{equation}
    t_E\left(R_E, D_l, \left|\boldsymbol{\mu}\right| \right) =  \frac{R_E}{\left|\boldsymbol{\mu}\right| D_l}\label{eqn:t_E} \; ,
\end{equation}
and is a function of the parameters $M_l$, $D_l$, $D_s$, and $\boldsymbol{\mu}$, since the Einstein ring is a function of $M_l$, $D_l$, and $D_s$. We compute $t_E$ for each of our simulated microlensing events in our ensemble directly from these parameters.

Since our goal is to use the microlensing planet frequency estimate of \citet{2010ApJ...720.1073G} to predict the number of planets RV surveys should find, we need to weight our microlensing parameters in some way that reflects this observational bias towards long timescale events. Because the exact functional form of their detection efficiency with respect to timescale is unknown (and likely unknowable), we adopt a simple function that reflects the overall trend of increasing efficiency with timescale. We assume that our simulated microlensing survey is not at all sensitive to events shorter than 10 days, and above 10 days, we are equally sensitive to all timescales. Our timescale efficiency function is then simply $\Phi\left(t_E\right) = \Theta\left(t_E-10\right)$, where $\Theta\left(t_E\right)$ is the Heaviside step function and $t_E$ is in units of days.

\begin{figure}[t!]
\epsscale{1.1}
\plotone{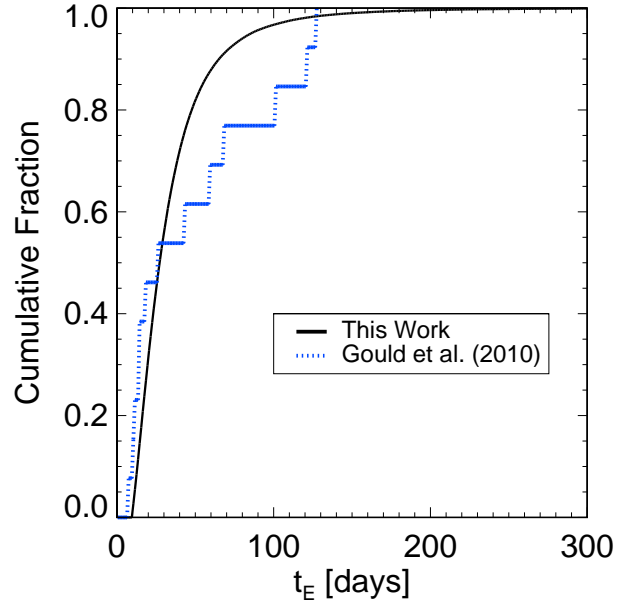}
\caption{The cumulative distribution function of event timescales obtained from our weighted distribution of timescales of our simulated microlensing events is shown by the black line. The blue, dotted line shows the CDF of the event timescales for the \citet{2010ApJ...720.1073G} microlensing sample. A two-sample K-S test fails to reject the null hypothesis that these samples are drawn from the same distribution.
  \label{fig:t_E_CDF_compare}}
\end{figure}

To include the effects of this selection function, we multiply the event rate by the value of this function to provide the statistical weights for each event. This gives all events with $t_E <10~$days zero weight. Figure \ref{fig:t_E_CDF_compare} displays our cumulative distribution function of $t_E$, comparing with that from the \citet{2010ApJ...720.1073G} sample. The median values of the two distributions are consistent (see \S~\ref{subsubsec:posterior_mlens_param_distributions}), and a two-sample K-S test yields a $D$-statistic of 0.26 with probability $P\left(D\right)=0.30$. We tried more complicated efficiency functions, but nothing reasonable provided better agreement.

\subsection{Generating an Ensemble of Microlensing Events}
\label{subsec:generating_an_ensemble_of_events}
We do not have analytic forms for the distributions of microlensing parameters, but we do know how the microlensing event rate depends on these parameters. Thus, we draw most of the event parameters from naive distributions and then weight each event by its event rate, given by equation (\ref{eqn:mlens_event_rate}), and a timescale bias.

\subsubsection{Drawing Event Parameters}
\label{subsubsec:drawing_params}
The priors we adopt for the microlensing event parameters are as follows. The source is assumed to always be in the bulge, at a distance of $D_s=10$~kpc. We choose 10~kpc because, as explained above, the source in all microlensing events is most probably be located between the center of the Galactic bulge and its back edge, which correspond to distances of 8 and 11~kpc, respectively, for the bulge model we adopt. Our results do not significantly change for fixed source distances within this range. The lens is allowed to either be in the bulge or in the disk (the way in which this is determined is described in \S~{\ref{subsubsec:computing_event_weights}}) and $D_l$ is drawn from a uniform distribution between $\left[0~{\rm kpc},10~{\rm kpc}\right]$. The lens star mass, $M_l$, is drawn from a distribution $d\Gamma/dM_l \propto M_l^{-1}$ between $\left[0.07~{\rm M_{\odot}},1.0~{\rm M_{\odot}}\right]$. Next, we draw two components of the lens-source relative proper motion in the North and East galactic directions, $\boldsymbol{\mu}=\left(\mu_N, \mu_E\right)$. Each component is drawn from a uniform distribution between $\left[0~{\rm mas~yr^{-1}}, 20.0~{\rm mas~yr^{-1}}\right]$. We show in figure~\ref{fig:mlens_param_dists} that there are no microlensing events with significant event rates that have relative proper motions $\left|\boldsymbol{\mu}\right|\gtrsim 15~{\rm mas~yr^{-1}}$. This is due to the exponential cutoff of high (and low) relative proper motions in the event rate, as shown in equation~(\ref{eqn:ls_rel_pm}). Thus, choosing components of the relative proper motion from a larger range, e.g. $\left[0~{\rm mas~yr^{-1}}, 200.0~{\rm mas~yr^{-1}}\right]$, results in exactly the same distribution.

The Galactic longitude and latitude of each microlensing event are drawn uniformly from within the intervals $-8^{\circ}\leq l \leq 0^{\circ}$ and $-5^{\circ}\leq b \leq 8^{\circ}$, respectively. These ranges in $l$ and $b$ cover the most dense fields monitored by the OGLE survey. We also select a time of observation for each microlensing event between the months of March and September (i.e. the``microlensing season,'' when the Galactic bulge is visible). This affects the event rate because it determines the velocity of the Earth along the line-of-sight of the microlensing event, $v_{\oplus}$, which determines the expected lens-source relative proper motion (see \S~\ref{subsubsec:lens_source_relative_proper_motion}). We later show that our results are insensitive to variations in the event location and the time of observation.

\subsubsection{Computing Relative Event Weights}
\label{subsubsec:computing_event_weights}
The weight for each simulated microlensing event is the product of its event rate and a timescale bias. We compute $R_E$, $v_{\rm rel}$, and $t_E$ for each simulated microlensing event using equations (\ref{eqn:einstein_ring_radius}), (\ref{eqn:v_rel}), and (\ref{eqn:t_E}). We need not worry about the mass term in the event rate equation because we draw our lens masses from what we assume to be an accurate mass function (i.e. $d\Gamma/dM_l\propto M_l^{-1}$), and thus each event is already weighted for the lens mass.

We compute the density and proper motion terms in the event rate equation as follows. Since the source is fixed in the bulge at $D_s=10$~kpc and the lens can be in either the bulge or the disk with $0~\text{kpc} < D_L < 10~$kpc, we compute the product of the density and proper motion terms, $\nu\left(x,y,z\right)[d\Gamma/d\boldsymbol{\mu}]$, as
\begin{align}
    & {} \nu\left(x,y,z\right)\frac{d\Gamma}{d\boldsymbol{\mu}} \propto \left[\rho_\text{l,disk}\left(x,y,z\right)\frac{d\Gamma_{disk}}{d\boldsymbol{\mu}}\right. \nonumber \\
    & {} \hspace{0.2in} + \left.\rho_\text{l,bulge}\left(x',y',z'\right)\frac{d\Gamma_{bulge}}{d\boldsymbol{\mu}}\right]\rho_\text{s,bulge}\left(x,y,z\right)\; , \label{eqn:nu_f_mu}
\end{align}
where $\rho_\text{disk}\left(x,y,z\right)$ is given by equation (\ref{eqn:disk_model}) and $\rho_\text{bulge}\left(x',y',z'\right)$ is given by equation (\ref{eqn:bulge_model}). The coordinates used in these models, $\left(x,y,z\right)$ and $\left(x',y',z'\right)$, depend on $D_l$, $l$, and $b$, and the appropriate transformations are described in section \S \ref{subsubsec:density_of_lenses}. For our purposes, we do not actually need to include the source term, $\rho_\text{s,bulge}\left(x,y,z\right)$, in equation (\ref{eqn:nu_f_mu}), since we keep the source distance, $D_s$, and the line-of-sight, $\left(l,b\right)$, of all our simulated microlensing events fixed, this source term is a constant for all events. However, if we were to account for variations in the source distance and event location, the source term would no longer be a constant for all events, and therefore we include it for completeness.

It is important to distinguish $d\Gamma_{disk}/d\boldsymbol{\mu}$ from $d\Gamma_{bulge}/d\boldsymbol{\mu}$ in equation (\ref{eqn:nu_f_mu}), as each requires different stellar velocities and velocity dispersions used in the computation of the event weights (see \S~{\ref{subsubsec:lens_source_relative_proper_motion}} for descriptions of the differing values for the disk and bulge). Since our disk has a 1~kpc hole centered on the galaxy, if $D_l$ and the line-of-sight are such that the lens lies within this hole, then the disk density, $\rho_\text{l,disk}$, is zero and the probability that the lens is in the disk is zero. Our bulge has a radius of 3~kpc centered on the galaxy, so if $D_l$ and the line-of-sight are such that the lens lies outside this range, then the bulge density, $\rho_\text{l,bulge}$, and the probability that the lens is in the bulge are zero. If $D_l$ and the line-of-sight place the lens within the limits of both the disk and the bulge, then both $\rho_\text{l,disk}\left(x,y,z\right)$ and $\rho_\text{l,bulge}\left(x',y',z'\right)$ are non-zero. In Bayesian terms, this can be described as
\begin{equation}
    P\left(D_l, D_s\right) \propto P\left(D_l|D_s\right)P\left(D_s\right)\; , \label{eqn:bayes_analog}
\end{equation}
where $P\left(D_l, D_s\right)=\rho\left(x,y,z\right)\frac{d\Gamma}{d\boldsymbol{\mu}}$, $P\left(D_l|D_s\right)=\rho_\text{l,disk}\left(x,y,z\right)[d\Gamma_{disk}/d\boldsymbol{\mu}] + \rho_\text{l,bulge}\left(x',y',z'\right)[d\Gamma_{bulge}/d\boldsymbol{\mu}]$ and $P\left(D_s\right)=\rho_\text{s,bulge}\left(x,y,z\right)$. We note that equation (\ref{eqn:bayes_analog}) would be an equality if we used the number density, $\nu\left(x,y,z\right)$, rather than the mass density, $\rho\left(x,y,z\right)$.

Finally, we multiply together the appropriate terms to calculate the weight, $W$, for each event,
\begin{equation}
    W = 2R_Ev_{\rm rel}\nu\left(x,y,z\right)\frac{d\Gamma}{d\boldsymbol{\mu}}\Phi\left(t_E\right)\; ,
\end{equation}
where the product $\nu\left(x,y,z\right)[d\Gamma/d\boldsymbol{\mu}]$ is given by equation (\ref{eqn:nu_f_mu}).

\subsubsection{Posterior Distributions of the Microlensing Event Parameters for our Simulated Sample}
\label{subsubsec:posterior_mlens_param_distributions}
\begin{figure*}
\epsscale{0.7}
\plotone{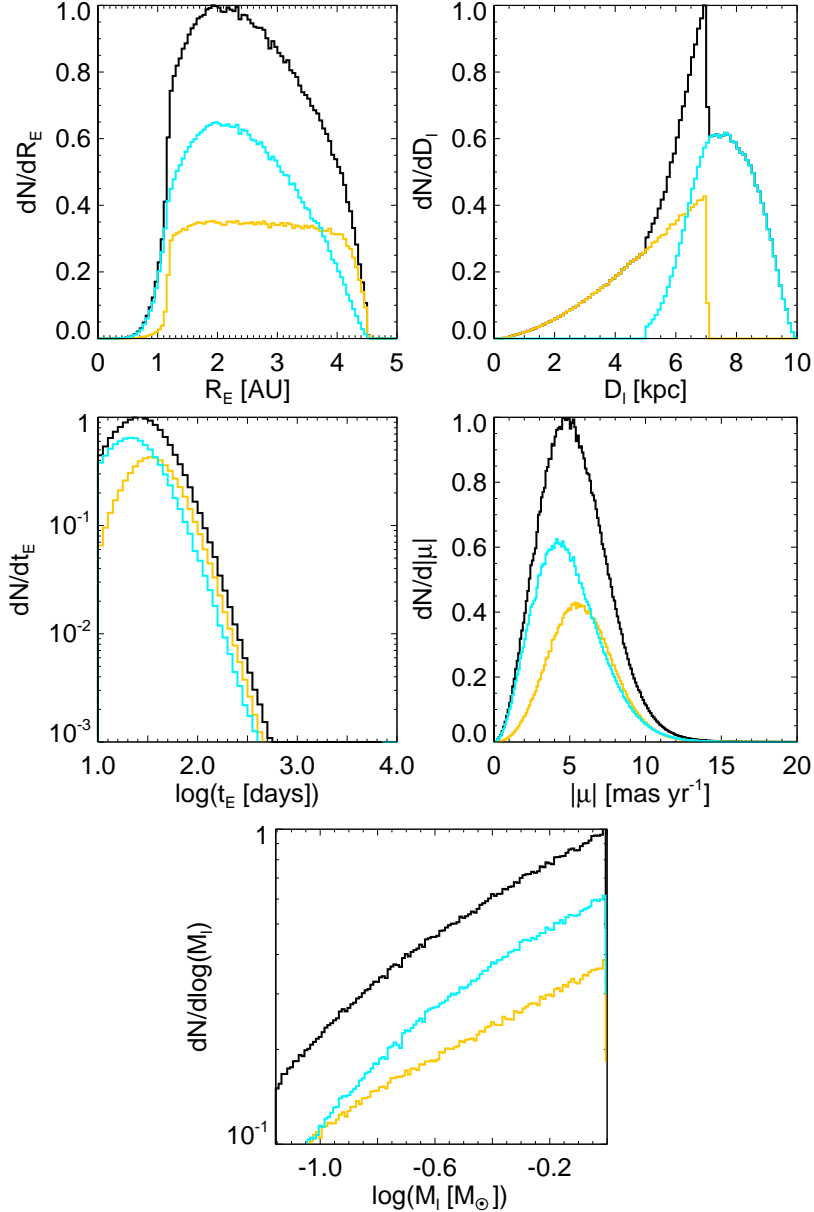}
\caption{Microlensing parameters for an ensemble of 10$^6$ simulated microlensing events. The blue line represents events where the lens star belongs to the bulge and the yellow line represents events where the lens star belongs to the disk. The black line represents all microlensing events, including both bulge and disk contributions. Each set of event parameters is weighted by their respective event rates and our timescale selection function. The ratio of bulge to total events is $\approx 59\%$.
  \label{fig:mlens_param_dists}}
\end{figure*}

We plot the appropriately weighted event parameters for our simulated microlensing events in figure \ref{fig:mlens_param_dists}. We find that the ratio of the number of events with a bulge lens to all events is roughly $59\%$ towards Baade's window, which agrees with the predictions of \citet{1994ApJ...430L.101K}, although we have adopted different density models for the disk and bulge and have included a bias in the timescale. Without the timescale bias (which \citet{1994ApJ...430L.101K} did not include), we still find a consistent fraction of bulge to disk events of about $63\%$. The weighted median values of the parameter distributions shown in figure \ref{fig:mlens_param_dists}, including contributions from both disk and bulge events, are $R_{E,{\rm med}}=2.52~$AU, $t_{E,{\rm med}}=27.9~$days, $D_{l,{\rm med}}=6.74~$kpc, $M_{l,{\rm med}} = 0.434~M_{\odot}$, and $\left|\boldsymbol{\mu}\right|_{\rm med}=5.12~{\rm mas~yr^{-1}}$. These values are in rough agreement with those of the subset microlensing events in the \citet{2010ApJ...720.1073G} sample for which they are measured, with the possible exception of the distribution of $D_l$. These events are listed in table \ref{tab:gould_events}. The median values for these events are $\left.R_{E,{\rm med}}\right|_{\rm G}=2.36~$AU, $\left.t_{E,{\rm med}}\right|_{\rm G}=30.5~$days, $\left.D_{l,{\rm med}}\right|_{\rm G}=3.35~$kpc, $\left.M_{l,{\rm med}}\right|_{\rm G} = 0.470~M_{\odot}$, and $\left|\boldsymbol{\mu}\right|_{\rm med, G}=4.70~{\rm mas~yr^{-1}}$. It should be noted that the comparisons of the median values of $D_l$ and $M_l$ are not exactly fair. Some of the events in the \citet{2010ApJ...720.1073G} sample for which these parameters are estimated have been inferred in a Bayesian sense, using a Galactic model and other priors, similar to our method of computing event rates for our simulated events, which then serve as the weights for our sample. We find that events in the \citet{2010ApJ...720.1073G} sample typically have smaller values of $D_l$ relative to our simulated sample.

\begin{table*}
\centering
\caption{\label{tab:gould_events} Best fit event parameters for the subset of published microlensing events included in the \citet{2010ApJ...720.1073G} sample with lens mass measurements, along with one unpublished event, OGLE-2007-BLG-349 (Dong et al., in preparation). We have used the reported parameters to compute the physical Einstein radius for each event as $R_E = \theta_ED_l$.}
\begin{tabular}{@{\extracolsep{0pt}}lcccccc@{\extracolsep{0pt}}}
\hline
\hline
Event Name & $t_E/{\rm days}$ & $M_l/M_{\odot}$ & $D_l/{\rm kpc}$ & $\left|\boldsymbol{\mu}\right|_{\rm geo}/{\rm mas~yr^{-1}}$ & $R_E/{\rm AU}$ & Reference$^{\ddagger}$ \\ \hline
\hline
OGLE-2007-BLG-224 & $6.91\pm 0.13$ & $0.056\pm 0.004$ & $0.525\pm 0.040$ & $48\pm 2$ & $\sim 0.48$ & (1) \\ \hline
OGLE-2008-BLG-279 & $101\pm 8$ & $0.64\pm0.10$ & $4.0\pm 0.6$ & $2.7\pm 0.2$ & $\sim 3.2$ & (2) \\ \hline
OGLE-2005-BLG-169 & $43\pm 4$ & $0.49^{+0.23}_{-0.29}$ & $2.7^{+1.6}_{-1.3}$ & $8.4\pm 1.7$ & $\sim 2.7$ & (3) \\ \hline
 MOA-2007-BLG-400 &  $14.3\pm 0.3$ & $0.30^{+0.19}_{-0.12}$ & $5.8^{+0.6}_{-0.8}$ & 8 & $\sim 1.9$ & (4) \\ \hline
OGLE-2007-BLG-349$^{\dagger}$ & 120 & $0.25\leq M_l/M_{\odot}\leq 0.45$ & $\sim 2.4$ & 3.1 & $\sim 2.5$ & (5) \\ \hline
OGLE-2007-BLG-050 &  68.09 & $0.50\pm 0.14$ & $5.47\pm 0.45$ & 2.58 & $\sim 2.6$ & (6) \\ \hline
 MOA-2008-BLG-310 &  11.14 & $\leq 0.67\pm 0.14$ & $>6.0$ & $5.1\pm 0.3$ & $\gtrsim 0.93$ & (7) \\ \hline
OGLE-2006-BLG-109 & 127.3 & $0.51^{+0.05}_{-0.04}$ & 1.49 & 4.3 & $\sim 2.2$ & (8) \\ \hline
OGLE-2005-BLG-188 &  14 & $0.16^{+0.21}_{-0.08}$ & $-$ & $-$ & $-$ & (9) \\ \hline
 MOA-2008-BLG-311 &  18 & $0.20^{+0.26}_{-0.09}$ & $-$ & $-$ & $-$ & (9) \\ \hline
\end{tabular}
\\
$^\dagger$The estimates of $D_l$ and $R_E$ for this event assume a lens mass of $M_l\sim 0.45~M_{\odot}$ (Dong et al., private communication). \\
$^\ddagger$(1) \citet{2009ApJ...698L.147G}; (2) \citet{2009ApJ...703.2082Y}; (3) \citet{2006ApJ...644L..37G}; (4) \citet{2009ApJ...698.1826D}; (5) Dong et al., in preparation; (6) \citet{batista09}; (7) \citet{2010ApJ...711..731J}; (8) \citet{2008Sci...319..927G,2010ApJ...713..837B}; (9) \citet{2010ApJ...720.1073G}.
\end{table*}

\subsection{Generating Planetary Parameters}
\label{subsec:generating_planetary_parameters}
Each microlensing event is independently given companion planets, described by two parameters: the projected separation, $s$, and the mass ratio $q$. We assume the joint distribution function of these parameters, $d^2N_{\rm pl}/(d\log{q}~d\log{s})$, is separable, such that $d^2N_{\rm pl}/(d\log{q}~d\log{s})=(dN_{\rm pl}/d\log{q})(dN_{\rm pl}/d\log{s})$. \citet{2010ApJ...720.1073G} found that the projected separations of their sample are consistent with being uniform in $\log{s}$, so we adopt a log-uniform distribution of projected separations for our simulated planet detections between $0.5\leq s \leq 2.5$, where $s$ is in units of the angular Einstein radius. The bounds on $s$ we adopt are determined in the following manner. \citet{2010ApJ...720.1073G} compute the absolute value of the log of the maximum detectable projected separation, $\left|\log{s_{\rm max}}\right|$, for each planet detection in their sample. These values range between $0.14 \leq \left|\log{s_{\rm max}}\right|\leq 0.55$. Since there are only six detected planets included in their sample, we adopt an intermediate value of $s\sim 2.5$ as the upper bound on our distribution of $s$ and a lower bound of $s\sim 0.5$, which is roughly the inverse of the upper bound (see \citet{1999A&A...349..108D} for information on the $s\leftrightarrow s^{-1}$ degeneracy).

Since \citet{2010ApJ...720.1073G} are unable to determine the mass ratio distribution, we adopt that of \citet{2010ApJ...710.1641S}, which has the form $dN_{\rm pl}/d\log{q} \propto q^{p}$ where $p=-0.68\pm 0.20$. We draw the mass ratios for each planet from the range $\left[10^{-5}, 10^{-2}\right]$ according to this distribution function, allowing the exponent to vary as a Gaussian with a mean value of $-0.68$ and standard deviation of $\sigma_p=0.20$. As we will show, variation in this exponent and variations in the normalization of our mass function are the main sources of (quantifiable) uncertainty in our calculations. We adopt a lower limit of $\log{q}=-5$ because this is where the planetary detection sensitivities of microlensing surveys begin to rapidly decline, although a more conservative value of $\log{q}=-4.5$ is typically used \citep{2010ApJ...720.1073G,2010ApJ...710.1641S}, and thus we are extrapolating the results slightly beyond their region of sensitivity. This extrapolation does not actually affect most of our results because, as we show in this paper and its companion paper, RV surveys are not sensitive to these planets (although it does affect the overall frequency estimate from the synthesized dataset in the companion paper).

\citet{2010ApJ...720.1073G} measured the frequency of planets in the mass ratio interval of $-4.5 < \log{q} < -2$. They found $d^2N_{\rm pl}/(d\log{q}~d\log{s}) \equiv\mathcal{G} =\left(0.36 \pm 0.15\right)~{\rm dex}^{-2}$ at the mean mass ratio $q_0 = 5\times 10^{-4}$. Because we use this frequency estimate of \citet{2010ApJ...720.1073G} but adopt the \citet{2010ApJ...710.1641S} mass ratio distribution, we must normalize our joint distribution function $d^2N_{\rm pl}/(d\log{q}~d\log{s})$ at the mean mass ratio and mean projected separation of the \citet{2010ApJ...720.1073G} sample in order to keep our simulated sample consistent with this measured frequency. This requirement is mathematically given by
\begin{align}
    & {} \mathcal{G} = \frac{1}{\Delta \log{s}\Delta \log{q}} \nonumber \\
    & {} \displaystyle \times \int_{\log{s_\text{min}}}^{\log{s_\text{max}}} \int_{\log{q_\text{min}}}^{\log{q_\text{max}}} \frac{d^2N_{\rm pl}}{d\log{q}~d\log{s}}d\log{q}~d\log{s}\; , \label{eqn:g_a_n}
\end{align}
where $\Delta\log{s} = \log{s_{\rm max}-\log{s_{\rm min}}}$ and $\Delta\log{q} = \log{q_{\rm max}-\log{q_{\rm min}}}$, and where $s_{\rm min}$, $s_{\rm max}$, $q_{\rm min}$, and $q_{\rm max}$ are the minimum and maximum values of the projected separation and mass ratio to which the \citet{2010ApJ...720.1073G} sample is sensitive, respectively. Our choice of priors yields
\begin{align}
    \frac{d^2N_{\rm pl}}{d\log{q}~d\log{s}} = & {} \frac{dN_{\rm pl}}{d\log{s}}\frac{dN_{\rm pl}}{d\log{q}} \nonumber \\
    = & {} \mathcal{A}\left(\frac{q}{q_0}\right)^{p}\; 
\end{align}
where $\mathcal{A}$ is the normalization factor we must adopt to force our priors to be consistent with the planet frequency found by microlensing. Thus, we have
\begin{align}
    & {} \left(0.36\pm 0.15\right)~\text{dex}^{-2} = \frac{1}{\Delta \log{q}} \nonumber \\
    & {} \hspace{0.6in} \displaystyle \times\int_{\log{q_\text{min}}}^{\log{q_\text{max}}} \mathcal{A}\left(\frac{q}{q_0}\right)^{-0.68\pm 0.20}d\log{q}\; , \label{eqn:a_G_n}
\end{align}
where $\log{q_\text{min}} = -4.5$ and $\log{q_\text{max}}=-2$, which yields a mean value and 68\% confidence interval of $\mathcal{A}=\left(0.23\pm 0.10\right)~{\rm dex^{-2}}$. We numerically compute the error on $\mathcal{A}$ by varying $p$ from a Gaussian distribution with mean value $-0.68$ and standard deviation $\sigma_p=0.20$, and by varying $\mathcal{G}$ according to a Poisson distribution. The latter is accomplished by drawing a non-integer value representing the average number of planets per star from a Poisson distribution with a mean value of 6 (i.e. the number of detections reported from the microlensing sample of \citealt{2010ApJ...720.1073G}) and calculating the implied planet frequency, $\mathcal{G}$, in a similar manner as described by \citet{2010ApJ...720.1073G}, but normalizing our results such that the mean value of $\mathcal{G}$ is equal to $0.36$.

Our resultant mass function is consistent with that derived independently by \citet{2012Natur.481..167C}, who find $d^2N_{\rm pl}/(d\log{a}~d\log{m_p})=\left(0.24^{+0.16}_{-0.095}\right)\left(m_p/M_{\rm Sat}\right)^{-0.73\pm 0.17}$. Although \citet{2012Natur.481..167C} write the distribution function in terms of $a$ and $m_p$, it is basically the same as our distribution function under our assumption of a characteristic lens mass of $M_l\sim 0.5~M_{\odot}$. \citet{2012Natur.481..167C} determined that the mass function is consistent with a flat distribution in $\log{a}$, which means it is also consistent with a flat distribution in $\log{s}$. However, we note that because we adopt a distribution function that is invariant in mass ratio rather than planet mass, the frequency of planets as a function of planet mass we predict depends on host star mass. For example, although our distribution function is consistent with \citet{2012Natur.481..167C} for $M_l\sim 0.5~M_{\odot}$, if we were to adopt a typical lens mass of $M_l\sim 0.3~M_{\odot}$ as assumed in \citet{2012Natur.481..167C}, we would infer a lower frequency at fixed planet mass of $d^2N_{\rm pl}/(d\log{a}~d\log{m_p}) = (0.16\pm 0.08)(m_p/M_{\rm Sat})^{-0.68\pm 0.20}$ (see \citet{clanton_gaudi14b} for a discussion of the uncertainty this introduces).

The total fraction of stars hosting planets in our simulated sample, which we designate by the constant $\mathcal{F}$, depends not only on the range of $\log{q}$ and $\log{s}$ we consider, but also the particular values of $p$ and $\mathcal{G}$. Mathematically, $\mathcal{F}_i$ is the effective ``area'' over which one of our simulated planetary microlensing events is sampled, weighted by the joint distribution function $d^2N_{\rm pl}/(d\log{q}~d\log{s})$, i.e.
\begin{equation}
    \mathcal{F} = \mathcal{A}\displaystyle \int_{\log{0.5}}^{\log{2.5}} \int_{-5}^{-2}\left(\frac{q}{q_0}\right)^{p}d\log{q}~d\log{s}\; . \label{eqn:f_a_n}
\end{equation}
We find a mean value and 68\% confidence interval of $\mathcal{F}=1.5\pm 0.6$, where we have calculated the uncertainties similar to the way we calculate them for $\mathcal{A}$. This is consistent with the conclusion from \citet{2012Natur.481..167C} that microlensing results imply roughly one planet per star.

\subsection{Marginalizing Over Microlensing Parameters and Planetary Mass Function Parameters}
\label{subsec:marginalizing_over_microlensing_parameters}
Because we do not know the exact scaling or normalization of the planetary mass function, we incorporate the uncertainties into our results in the following manner. Consider one realization $i$ where the planetary mass function is given by $d^2N_{\rm pl}/(d\log{q}~d\log{s}) = \mathcal{A}_i\left(q/q_0\right)^{p_i}$, where $\mathcal{A}_i$ is a function of the values of $\mathcal{G}_i$ and $p_i$ given implicitly by equation~(\ref{eqn:g_a_n}), we compute the implied fraction of stars hosting planets, $\mathcal{F}_i$, via equation~(\ref{eqn:f_a_n}) with $\mathcal{A}\rightarrow\mathcal{A}_i$ and $p\rightarrow p_i$. We then generate an ensemble of $N_{\rm ml}$ microlensing events due to planetary systems in this realization and compute their respective weights, treating the events as we did in \S~\ref{sec:typical_dists} and producing a joint distribution of $K$ and $P$ for each. We then map the parameters of each planetary microlensing event into a single $(K, P)$ pair according to the methods presented in \S~\ref{subsec:eccentric_orbits}. Each of these individual distributions, $\left[d^6N_{\rm pl}/(dKdPdM_ldD_ld\log{q}d\log{s})\right]_j$, formally given by equation (\ref{eqn:full_mlens_dist}), is normalized such that the total integrated probability is equal to the weight, $W_j$, of the particular event (refer to \S~\ref{subsubsec:computing_event_weights}), i.e.
\begin{equation}
    W_j = \displaystyle \int \int \left.\frac{d^6N_{\rm pl}}{dKdPdM_ldD_ld\log{q}d\log{s}}\right|_jdKdP\; .
\end{equation}
We then sum up all these individual distributions, normalizing the resultant distribution by the total of the weights of all the microlensing events in this realization, and multiplying by the total fraction of stars hosting planets with properties in the sampled region of parameter space (i.e. log mass ratios in the range $-5 \leq \log{a} \leq -2$ and projected separations in the range $0.5\leq s \leq 2.5$) in this realization, $\mathcal{F}_i$, which yields
\begin{equation}
    \left.\frac{d^2N}{dKdP}\right|_{i} = \mathcal{F}_i\frac{\displaystyle \sum_j \left.\frac{d^6N_{\rm pl}}{dKdPdM_ldD_ld\log{q}d\log{s}}\right|_j}{\displaystyle \sum_j W_j}\; .
\end{equation}
This is the distribution of the RV observables $K$ and $P$ for analogs to the entire population of planets inferred from microlensing in this realization. 

We repeat this procedure for many different realizations, each corresponding to a different planetary mass function consistent with the measurements by \citet{2010ApJ...720.1073G} and \cite{2010ApJ...720.1073G}. We draw the values of $p_i$ for each realization from a Gaussian with a mean value of $-0.68$ and a standard deviation of $\sigma_p=0.20$, while we vary $\mathcal{G}_i$ according to a Poisson distribution. Specifically, to determine the distribution of $\mathcal{G}$ values, we draw an integer number of planet detections from a Poisson distribution with a mean value of 6 (i.e. the number of detections reported from the microlensing sample of \citealt{2010ApJ...720.1073G}) and calculate the implied planet frequency, $\mathcal{G}_i$, in a similar manner as described by \citet{2010ApJ...720.1073G}. Our final result is then the mean of the distributions for all the realizations and we compute confidence limits from variations in these distributions due to variations in the values of $\mathcal{G}_i$ and $p_i$ for each realization.

\section{Results}
\label{sec:results}
\subsection{Mapping the Entire Microlensing Parameter Space to the RV Observables $K$ and $P$}
\label{subsec:mapping_mlens_to_RV}
\begin{figure*}
\epsscale{0.9}
\plotone{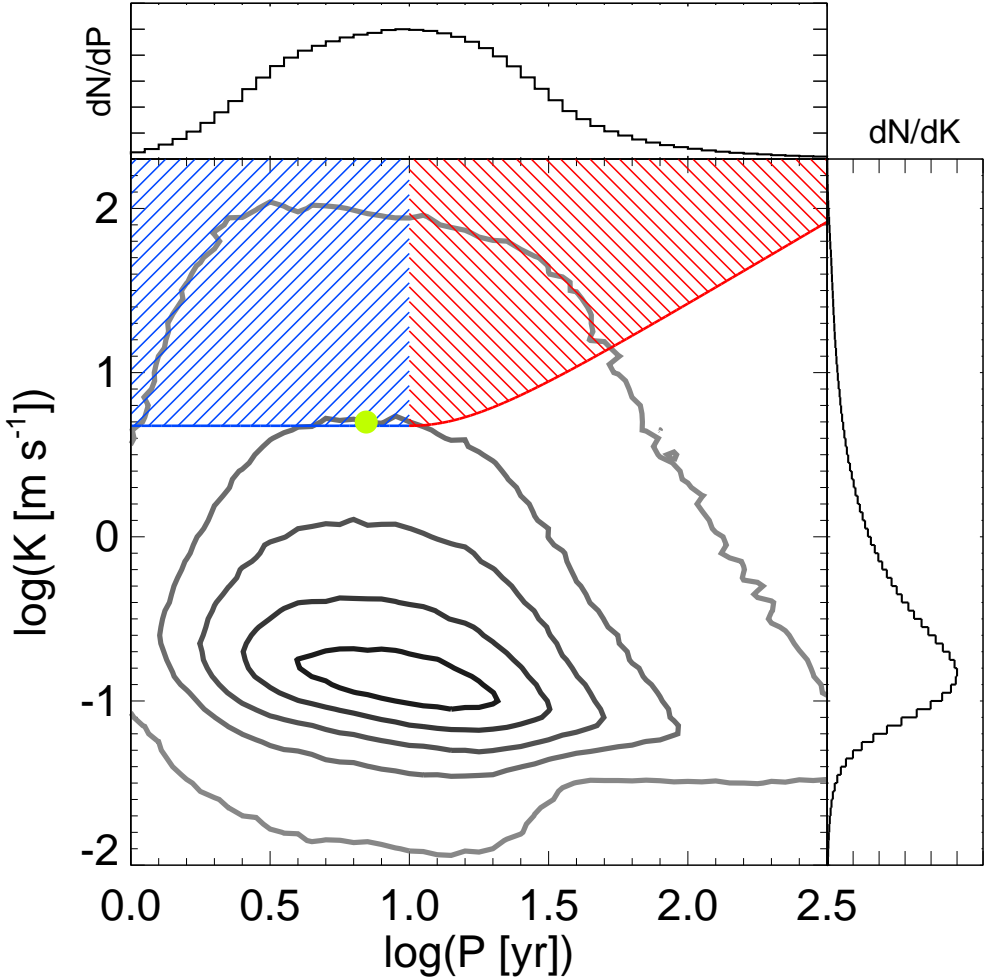}
\caption{Contours of the probability density of $K$ and $P$, marginalized over the entire microlensing parameter space, are shown in greyscale. The contour levels, going from grey to black, are $1\%$, $10\%$, $25\%$, $50\%$ and $80\%$ of the peak density. The filled yellow circle represents where the typical microlensing planet lies in this parameter space at the median inclination and mean anomaly and on a circular orbit ($K_{\rm typ}\sim 5~{\rm m~s^{-1}}$, $P_{\rm typ}\sim 7~$yr). The median value of the period distribution is $\approx 9.4~$yr and the median value of the velocity semi-amplitude distribution is $\approx 0.24~{\rm m~s^{-1}}$. The 68\% intervals in $K$ and $P$ are $0.0944\leq K/{\rm m~s^{-1}}\leq 1.33$ and $3.35\leq P/{\rm yr}\leq 23.7$, respectively. The blue and red colored line represents a constant SNR of $\mathcal{Q}_{\rm min}=5$, the form of which is given by equation (\ref{eqn:snr_full}). The blue shaded area represents the region of parameter space where planets are detected at $\mathcal{Q}\geq 5$, and have $P\leq 10~$yr. The red shaded area represents the region of parameter space where long-term trends are detected by our fiducial RV survey ($N\approx 30$, $\sigma \approx 4~{\rm m~s^{-1}}$, $T\approx 10~$yr). For $\mathcal{Q}_{\rm min}=5$, corresponding to a detection efficiency given by equation (\ref{eqn:sn_5_det_eff}), we predict that such a survey should (on average) detect $f_{\rm det}= 0.049^{+0.046}_{-0.026}$ planets per star and see $f_{\rm tr} = 0.024^{+0.024}_{-0.014}$ trends per star due to planets.
  \label{fig:marg_dist}}
\end{figure*}

The final, marginalized distribution of RV observables, shown in figure \ref{fig:marg_dist}, represents a full microlensing parameter space mapping into the radial velocity values $K$ and $P$. This plot is the median joint distribution of $K$ and $P$ resulting from varying planetary mass functions consistent with the observations of \citet{2010ApJ...720.1073G} and \citet{2010ApJ...710.1641S}, each of which includes an ensemble of microlensing events observed along a single line-of-sight towards Baade's Window ($l=-1^{\circ}$, $b=-3.9^{\circ}$) at a single epoch. The peak of the $K$ distribution, when marginalized over all $P$, is set by the lower bound on the value of $q$ we sample from our prior of the mass ratio distribution. If we had set the lower bound of our prior on $q$ lower, this peak would occur at a yet lower value of $K$. However, because we choose our normalization of each planetary mass function such that there is a fixed number of planets per dex$^{2}$, continuing the distribution to lower values of $q$ does not affect our estimate of the number of detectable (or trending) planets, provided that the minimum $q$ is set so that the resulting $K$ and $P$ are always well outside the region of detectability. The spread in the $P$ distribution, when marginalized over $K$, is set primarily by the limits of $s$ where we sample our prior on the projected separation and the distribution of eccentricities.

The median values of this joint distribution of $K$ and $P$ are $\approx 0.24~{\rm m~s^{-1}}$ and $\approx 9.4~$yr, respectively. The 68\% intervals in $K$ and $P$ are $0.0944\leq K/{\rm m~s^{-1}}\leq 1.33$ and $3.35\leq P/{\rm yr}\leq 23.7$, respectively, and the 95\% intervals are $0.0422\leq K/{\rm m~s^{-1}}\leq 16.8$ and $1.50\leq P/{\rm yr}\leq 94.4$, respectively. Although the median value of the $K$ distribution is sensitive to our choice of the lower limit on the mass ratio, $\log{q}=-5$, this lower limit is not arbitrary, as it roughly corresponds to the cutoff in the planetary detection sensitivity by microlensing surveys \citep{2010ApJ...720.1073G,2010ApJ...710.1641S}. Thus, $K\approx 0.24~{\rm m~s^{-1}}$ then represents the median value of $K$ at which RV surveys would detect planets if they were sensitive to all the planets for which microlensing surveys are sensitive. This median $K$ differs significantly from our order of magnitude estimate in \S~\ref{sec:OoM}, which was $\sim 5~{\rm m~s^{-1}}$ for the typical microlensing planet. The difference between the median value found by marginalizing over the microlensing parameter space and the order of magnitude estimate arises from the steeply declining planetary mass function inferred from microlensing surveys and because we consider somewhat lower mass ratios than the minimum value of the \citet{2010ApJ...720.1073G} sample.

We ran similar simulations where the line-of-sight was allowed to vary within the most dense fields monitored by OGLE, where $l$ and $b$ were chosen uniformly between $[-8^{\circ},0^{\circ}]$ and $[-5^{\circ},8^{\circ}]$, respectively. We also ran simulations where the time of observation was allowed to vary between dates during the ``microlensing season,'' when the bulge is visible between March and September. Allowing for different observation times changes the velocity of the Earth along the line-of-sight, $v_{\oplus}$, which affects the expected lens-source relative proper motion (see \S \ref{subsubsec:lens_source_relative_proper_motion}). We find that variations in the line-of-sight and the Earth's velocity along the line-of-sight do not significantly change our results. The median of the resultant $K$ and $P$ distributions only change by as much as $\sim 0.003~{\rm m~s^{-1}}$ and $\sim 0.07~$yr, respectively, from the case where the line-of-sight is fixed towards Baade's Window and the Earth's velocity is constant along this line-of-sight for all microlensing events. For simplicity, and since our stellar density models are normalized along the line-of-sight towards Baade's window (see \S~\ref{subsubsec:density_of_lenses}), we fix our line-of-sight to be that towards Baade's Window and Earth's velocity along this line-of-sight to be a constant for all simulations in this study.

\subsection{Estimate of the Fraction of Analogs to Those Planets Detected by Microlensing Observable by RV Surveys}
\label{subsec:estimate_percent_mlens_detected}
We can use our joint $K$ and $P$ distribution, marginalized over all microlensing parameter space and shown in figure \ref{fig:marg_dist}, to predict an estimate of the fraction of microlensing planets that are detectable by RV surveys. In order to do this, we need a rough estimate of the detection efficiency for RV surveys, i.e. a minimum mass to which they are sensitive as a function of orbital period. We estimate this by solving equation (\ref{eqn:snr_full}) for $K$ and then transforming this $K$ into $m_p\sin{i}$ via the velocity semi-amplitude equation for a circular orbit. This yields $m_p\sin{i}$ as a function of the minimum SNR, which we designate as $\mathcal{Q}_{\rm min}$ for the predicted detections:
\begin{align}
    \left.m_p\sin{i}\right|_{\rm min} = & {} \sqrt{\frac{2}{N}}\sigma\left\{1-\frac{1}{\pi^2}\left(\frac{P}{T}\right)^2\sin^2{\left(\frac{2\pi T}{P}\right)}\right\}^{-1/2} \nonumber \\
    & {} \times M_l^{2/3}\left(\frac{P}{2\pi G}\right)^{1/3}\mathcal{Q}_{\rm min}\; . \label{eqn:mpsini_sn}
\end{align}
Substituting in the values for the number of epochs per star and average precision for our fiducial RV survey of $N\approx 30$ and $\sigma \approx 4~{\rm m~s^{-1}}$, respectively, and assuming $P\ll T$, we estimate the minimum $m_p\sin{i}$ as a function of orbital period to which our fiducial RV survey is sensitive to be
\begin{align}
    \left.m_p\sin{i}\right|_{\rm min} \approx & {} \; 69~{\rm M_{\oplus}}\left(\frac{P}{7~{\rm yr}}\right)^{1/3} \nonumber \\
    & {} \; \times \left(\frac{M_l}{0.5~{\rm M_{\odot}}}\right)^{2/3} \left(\frac{\mathcal{Q}_{\rm min}}{5}\right)\label{eqn:sn_5_det_eff}
\end{align}
for periods less than the time baseline of observations. This is still likely to be in the giant planet regime (i.e. $m_p \gtrsim 0.1~M_{\rm Jup}$). For periods larger than this baseline, RV surveys would only see a fraction of the period of the planetary orbit and so could not claim a detection at this SNR. However, these planets would be seen as RV drifts, or trends, the sensitivity of which would be given by equation (\ref{eqn:mpsini_sn}) for the chosen values of $N$, $\sigma$ and $T$. We also assume that the full range of eccentricities can be identified by such RV surveys as detections or trends. In reality, significant selection effects hinder detection of planets with orbital eccentricities $e\gtrsim 0.6$ \citep{2004MNRAS.354.1165C}. However, due to our choice of eccentricity prior, only about 8\% of planets have $e\geq 0.6$. While this introduces uncertainty to our estimate of detectable planets and trends, the total error is dominated by the uncertainty in the quantity $\mathcal{F}$, arising from uncertainties in both the normalization and slope of the planetary mass function as measured by \citet{2010ApJ...720.1073G} and \citet{2010ApJ...710.1641S}, respectively.

We use the estimate of detection sensitivity given by equation~(\ref{eqn:sn_5_det_eff}) to predict the fraction of analogs to the planets inferred from microlensing surveys that RV surveys should detect as planets or long-term drifts by applying a SNR cut in the $\left(K,P\right)$ parameter space of the microlensing population and count the number of relative planets that lie in the ``detectable'' and ``trending'' regions. Figure \ref{fig:marg_dist} illustrates this procedure. The predicted number of planets per star that will be detected by such an RV survey, which we designate as $f_{\rm det}$, is calculated as
\begin{equation}
    f_{\rm det} = \displaystyle  \frac{1}{N_{\rm ml}} \sum_i \mathcal{F}_i N_{\rm det, i}\; , \label{eqn:f_det_n}
\end{equation}
where $N_{\rm ml}$ is the number of microlensing events per realization (see \S~\ref{subsec:marginalizing_over_microlensing_parameters}), $N_{\rm det, i}$ is the number of planets that are detectable according to equation~(\ref{eqn:sn_5_det_eff}) and thus lie in the blue region of figure~\ref{fig:marg_dist}, and $\mathcal{F}_i$ is the fraction of stars hosting planets in a given realization. Similarly, the number of planets per star we predict to be seen as trends by our fiducial RV survey, which we designate as $f_{\rm tr}$, is given by equation~(\ref{eqn:f_det_n}) but with $N_{\rm det, i}\rightarrow N_{\rm tr, i}$, where $N_{\rm tr, i}$ is the number of planets that lie in the red region of figure~\ref{fig:marg_dist}, i.e. the number of planets that produce detectable radial velocities but have orbital periods longer than the duration of our ficucial RV survey.

In the mathematical formalism developed in \S~\ref{sec:typical_dists}, the predicted number of planets per star is given by $f_{\rm det} = N_{\rm det}/N_{\star}$, where $N_{\star}$ is the number of stars monitored by a given RV survey and where $N_{\rm det}$ is given by equation~(\ref{eqn:n_pl_obs}), which takes the form
\begin{equation}
    N_{\rm det} = \displaystyle \int \int \frac{d^2N}{dKdP}\Phi\left(K\right)\Phi\left(P\right)dKdP\; , \label{eqn:n_det_math}
\end{equation}
where $\Phi\left(K\right) = \Theta\left(K-K_{\rm det}\right)$ and $\Phi\left(P\right) = \Theta\left(T-P\right)$ are the selection functions that determine which planets are detectable. Here, $K_{\rm det}=K\left(\mathcal{Q}_{\rm min}, N, \sigma, T, P\right)$ is the minimum detectable $K$. The functional form of $K_{\rm det}$ is found by solving equation (\ref{eqn:snr_full}) for $K$ and substituting in the appropriate parameter values. The mathematical expression for the expected number of trends is the equivalent to equation (\ref{eqn:n_det_math}) except in this case, we have $\Phi\left(P\right) = \Theta\left(P-T\right)$.

For a SNR cut of 5 or higher (i.e. $\mathcal{Q}_{\rm min}=5$), corresponding to a detection efficiency given by equation (\ref{eqn:sn_5_det_eff}), we predict that RV surveys should on average detect $f_{\rm det}= 0.049^{+0.046}_{-0.026}$ planets per monitored star and find $f_{\rm tr}= 0.024^{+0.024}_{-0.014}$ long-term RV trends per monitored star resulting from planets. The uncertainties in these quantities come from the error in $\mathcal{F}$, which arise from the uncertainties in the normalization and slope of the planetary mass function from the measurements of $\mathcal{G}$ and $p$ by \citet{2010ApJ...720.1073G} and \citet{2010ApJ...710.1641S}, respectively (refer to \S~\ref{subsec:generating_planetary_parameters} for details). Thus, if an RV survey were to monitor a sample of $N_{\star}=100$ M dwarfs with a mass distribution covering the interval $0.07\leq M_{\star}/M_{\odot}\leq 1.0$ uniformly in $\log{M_{\star}}$, we predict should should detect $N_{\rm det} = f_{\rm det}N_{\star} = 4.9^{+4.6}_{-2.6}$ giant planets and identify $N_{\rm tr}= f_{\rm tr}N_{\star}= 2.4^{+2.4}_{-1.4}$ trends due to a giant planet, at a minimum SNR of 5. If we relax the SNR constraint to $\mathcal{Q}_{\rm min}= 1$, which would correspond to a sensitivity of
\begin{align}
    \left.m_{\rm p}\sin{i}\right|_{\rm min} \approx & {} \; 14~{\rm M_{\oplus}}\left(\frac{P}{7~{\rm yr}}\right)^{1/3} \nonumber \\
    & {} \; \times \left(\frac{\mathcal{Q}_{\rm min}}{1}\right)\left(\frac{M_l}{0.5~{\rm M_{\odot}}}\right)^{2/3}\; \label{eqn:sn_1_det_eff}
\end{align}
for periods less than 10~yr, we find $f_{\rm det}=0.16^{+0.10}_{-0.06}$ and $f_{\rm tr}=0.082^{+0.062}_{-0.037}$. This would mean our fiducial RV survey should detect $N_{\rm det}= 16^{+10}_{-6}$ planets and see $N_{\rm tr}= 8.2^{+6.2}_{-3.7}$ trends due to planets from a sample of 100 M dwarfs with a mass distribution covering the interval $0.07\leq M_{\star}/M_{\odot}\leq 1.0$ uniformly in $\log{M_{\star}}$.

Of course, in real RV surveys, the number of epochs, time baselines and RV precisions are different for each star in their sample. This means that each star has different detection sensitivities, so while the above results give a rough estimate of the fraction of predicted detections and trends, a more careful analysis is needed to more accurately compare microlensing predictions with current RV studies. This is the aim of an upcoming paper in which we will apply the methodology presented in this work to derive distributions of RV observables for analogs of the planets found by microlensing. After carefully estimating RV detection sensitivies of current RV surveys of M dwarf samples, we will predict the number of planets these surveys should find and compare with the number of planet detections they report.

\section{Discussion}
\label{sec:discussion}
Our order of magnitude calculation demonstrates that an RV survey with an average number of epochs per star $N=30$, average precision $\sigma=4~{\rm m~s^{-1}}$, and time baseline $T=10~$yr, should be able to detect the typical microlensing planet ($M_l\sim0.5~M_{\odot}$, $M_p\approx 0.26~M_{\rm Jup}$, $r_{\perp}\sim 2.5~{\rm AU}$) on a circular orbit with an inclination equal to the median value of a uniform prior in $\cos{i}$ at a SNR of about 5. However, if we allow the orbital parameters of this typical microlensing-detected planet to vary, we find that it is not always detectable via RVs. This motivates our study to map analogs of the population of planets inferred from microlensing into the RV observables $K$ and $P$ to determine the number of planets per star RV surveys should detect and see as trends. In doing so, we are able to show that although the regions of planet parameter space where microlensing and RV surveys are sensitive are largely disjoint, there is some overlap for giant planets with masses $m_p\gtrsim 10^2~M_{\oplus}$ and with orbital periods between $\sim 3-10~$years.

An RV survey with a duration of $\sim 10~$years has significant overlap with microlensing surveys in terms of period. We find that the median period of analogs to the planets inferred by microlensing is $P_{\rm med}\approx 9.4~$years, with a 68\% interval of $3.35\leq P/{\rm yr}\leq 23.7$. However, due primarily to a steeply declining planetary mass function, we find that the median velocity semi-amplitude of such planets is $K_{\rm med}\approx 0.24~{\rm m~s^{-1}}$, with a 68\% interval of $0.0944\leq K/{\rm m~s^{-1}}\leq 1.33$. The steeply declining planetary mass function measured by microlensing is primarily responsible for the shape of the posterior distribution of $K$ for these planets, and puts a majority of the planet population inferred from microlensing out of reach for RV surveys. The detection sensitivity of RV surveys to such planets is thus ultimately limited by their velocity precision. The best precisions of state-of-the-art RV surveys is $\sim$few ${\rm m~s^{-1}}$  (including both instrumental errors as well as stellar jitter), so the bulk of the the planet population inferred from microlensing is inaccessible by RVs. In order for RV surveys to access the bulk of these planets, precisions of $\sim 0.1~{\rm m~s^{-1}}$ need to be achieved. However, in order to detect the large population of giant planets ($m_p\gtrsim 0.1~M_{\rm Jup}$) inferred from microlensing, precisions of only $\sim 1~{\rm m~s^{-1}}$ are needed.

We find that orbital eccentricity and inclination also significantly impact the mapping of microlensing parameters into the RV observables $K$ and $P$, although to a lesser extent than the planetary mass function. Allowing for eccentric orbits greatly increases the area of allowed parameter space (see figures \ref{fig:kp_contours_typical} and \ref{fig:kp_contours_ecc_test}) and variations in the inclination increase the spread of $K$ values directly, since $K \propto \sin{i}$ for a given period, host star mass and mass ratio. Nonetheless, we find that variation of the orbital parameters tend not to shift the median $K$ and $P$, but mostly serve to increase the spread in these observables.

Despite the fact that RV and microlensing surveys are sensitive to different parameters, have very different targeting/observing strategies, and thus suffer from different selection effects, we have demonstrated that it is possible to statistically compare detection results from these two completely independent techniques. We find that a RV survey monitoring a sample of 100 M dwarfs with a mass distribution covering the interval $0.07\leq M_{\star}/M_{\odot}\leq 1.0$ uniformly in $\log{M_{\star}}$, and with an average number of epochs per star $N=30$, average sensitivity $\sigma=4~{\rm m~s^{-1}}$, and time baseline $T=10~$yr, should on average detect $4.9^{+4.6}_{-2.6}$ giant planets and identify $2.4^{+2.4}_{-1.4}$ as long-term trends at a SNR of 5 or higher if analogs of the planets detected by microlensing orbit local M dwarfs. In practice, real RV surveys could have different sensitivities than those implied by our fiducial survey. A careful comparison of microlensing and RV detection results will require knowledge of the specific sensitivities of all stars monitored by RVs, but the methodology of comparison will be very similar to that laid out in this study. In a companion paper, we do just this, comparing microlensing results with the M dwarf RV studies of \citet{2010PASP..122..905J} and \citet{2013A&A...549A.109B}.

\acknowledgments
This research has made use of NASA's Astrophysics Data System and was partially supported by NSF CAREER Grant AST-1056524. We thank John Johnson and Benjamin Montet for helpful comments and conversations.

\bibliographystyle{hapj}
\bibliography{myrefs}
\end{document}